# A survey on Big Data and Machine Learning for Chemistry

Jose F Rodrigues Jr[1], Larisa Florea[3], Maria C F de Oliveira[1],
Dermot Diamond[3], Osvaldo N Oliveira Jr[2]

[1]Institute of Mathematical Sciences and Computing, University of São Paulo (USP),
[2]São Carlos Institute of Physics, University of São Paulo (USP)
13560-970 São Carlos, SP, Brazil

[3]Dublin City University (DCU)
Glasnevin, Dublin 9, Dublin, Ireland

[junio, cristina]@icmc.usp.br, larisa.florea@dcu.ie, dermot.diamond@dcu.ie, chu@ifsc.usp.br

**Outline**

1. Introduction

2. The new trend with big data and machine learning

    2.1 Big data

    2.2 Machine learning

    2.3 An overview of Deep Learning

    2.4 The flow path towards data-based scientific discovery in chemistry

3. Materials Discovery

    3.1 Large databases and initiatives

    3.2 Identification of compounds with genetic algorithms

    3.3 Synthesis prediction using machine learning

    3.4 Quantum chemistry

    3.5 Computer-aided drug design

4. Sensors and biosensors for intelligent systems and internet of things

    4.1 Big data methods and ML in data analysis

    4.2 Providing data for big data and ML applications with chem/bio sensor networks

    4.3 Prospects for scalable applications of chemical sensors and biosensors

5. Current limitations and future outlook

**Abstract**

Herein we review aspects of leading-edge research and innovation in chemistry which exploits big data and machine learning (ML), two computer science fields that combine to yield machine intelligence. ML can accelerate the solution of intricate chemical problems and even solve problems that otherwise would not be tractable. But the potential benefits of ML come at the cost of big data production; that is, the algorithms, in order to learn, demand large volumes of data of various natures and from different sources, from materials properties to sensor data. In the survey, we propose a roadmap for future developments, with emphasis on materials discovery and chemical sensing, and within the context of the Internet of Things (IoT), both prominent research fields for ML in the context of big data. In addition to providing an overview of recent advances, we elaborate upon the conceptual and practical limitations of big data and ML applied to chemistry, outlining processes, discussing pitfalls, and reviewing cases of success and failure.

**1. Introduction**

The ongoing revolution with artificial intelligence is bound to transform society, well beyond applications in science and technology. A key ingredient is machine learning (ML) for which increasingly sophisticated methods have been developed, thus bringing an expectation that within a few decades, machines may be able to outperform humans in most tasks, including intellectual ones. Two converging movements are responsible for the revolution. The first may be referred to as "data-intensive discovery", "e-Science", or "big data" [1,2], a movement wherein massive amounts of data – including valuable information – are transformed into knowledge. This is attained with various computational methods, which increasingly engage ML techniques, in a movement characterized by a transition in which data moves from a "passive" to an "active" role. What we mean by an "active" role for data is that it is not considered solely to confirm or refute a hypothesis, but also to assist in raising new hypotheses to be tested, at an unprecedented scale. The transition into a fully active role for data will only be complete when the computational methods (or machines) are capable of generating knowledge themselves (see below). Within this novel paradigm, data must be organized in such a way as to be machine-readable, particularly since computers at present cannot "read" and interpret. Attempts to teach computers to read are precisely within the realms of the second movement, in which natural language processing tools are under development to process spoken and written text. Significant advances in this regard have been recently achieved upon combining ML and big data, as may be appreciated by the astounding progress in speech processing [3] and machine translation. Computational systems are still far away from the human ability to interpret text, but the increasingly synergistic use of big data and ML allows one to envisage the creation of intelligent systems that can handle massive amounts of data with analytical ability. Then, beyond the potential to outperform humans, machines would also be able to generate knowledge – without human intervention.

Regardless of whether such optimistic predictions will become a reality, big data and ML already have a significant impact owing to the generality of their approaches and uses. To understand why this is happening, we need to distinguish the contributions of the two areas. Working with big data normally requires proper infrastructure, with major difficulties being associated with the gathering and curation of a lot of data. In chemistry, for instance, access to large databases and considerable computational power are essential, as exemplified in this review paper in the discussion of sensor networks. Standard ML algorithms, on the other hand, can operate on small datasets and in many cases require only limited computational resources. Furthermore, there are major limitations in terms of what ML can achieve owing to fundamental conceptual difficulties.

The goals of ML fall into two distinct types [4]: i) classification of data instances in a large database, as in image processing and voice recognition; ii) making inferences based on the organization and/or structuring of the data. Needless to say, the second goal is much harder to achieve. Let us consider, as an example, the application of ML to identify text authorship. In a classification experiment with tens of English literature books modelled as word networks, book authors were identified with high accuracy using supervised ML [5]. Nonetheless, it would be impossible with current technology to make a detailed analysis of writing style and establish correlations among authors, which would represent a task of the second type. This will require considerable new developments, such as teaching computers to read. Today, the success of ML stems mostly from applications focused on the first goal, which encompasses most of its applications in chemistry. Nonetheless, a lot more can be expected in the next few decades, as we shall comment upon in our Outlook section.

Considerable work has been devoted to addressing the challenges that arise when chemistry meets big data and ML. The evolution in computing resources allows chemists to produce and manipulate unprecedented data volumes to be stored and managed via algorithms with embedded intelligence [6]. Data processing has become a much more complex endeavour than just storage and retrieval, as concepts such as data curation and provenance come into play, particularly if the data is to be machine-readable. Chemists already employ substantial amounts of machine-readable data in at least two major fields: in exploring protein databanks in crystallography. These are illustrative examples of machine-readable content that require little (if any) artificial intelligence.

In this review paper, we chose two particular areas in chemistry to illustrate our points: sensing, which is already becoming ubiquitous; and materials discovery based on physicochemical properties. Big data and ML serve materials science, for example, through materials discovery and data analysis. However, at the same time, data and ML are served by materials science, for example, as data-intensive new sensors and devices are created for new applications related to the Internet of Things (IoT).

## 2. New trends in big data and machine learning relevant to chemistry

Concepts and methodologies related to big data and ML have been employed to address many problems in chemistry over the years, as emphasized by the illustrative examples presented in Sections 3 and 4. A description of the concepts and myths of big data and machine learning targeted at chemists and related professionals is given in the review by Richardson et al. [7]. In this section, we shall briefly introduce such concepts as background to assist the reader to follow the remaining sections. Throughout the text, we shall use the acronyms **AI** for *Artificial Intelligence*, **ML** for *Machine Learning*, **DL** for *Deep Learning*, and **DNN** for *Deep Neural Networks* (the latter two used interchangeably as synonyms).

### 2.1 Big data

The broadly advertised term "big data" has gained widespread attention as a direct consequence of the rapid growth in the amount of data being produced in all fields of human activity. The magnitude of this increase is often highlighted even to the general public, as in a recent news piece by Forbes, which states that "There are 2.5 quintillion bytes of data created each day at our current pace, but that pace is only accelerating with the growth of the Internet of Things (IoT)". [8] But the term *big data* is not just about massive data production, a perspective that has popularized it as a jargon filled with expectations. Big data also refers to a collection of novel software tools and analytical techniques that can generate real value by identifying and visualising patterns and trends from disperse and apparently unconnected data sources. Nevertheless, in order to grasp the genuine virtues and potential of big data, its meaning must be interpreted in connection with the specificities of a particular domain or purpose, as in the case of chemistry. Big data might be understood as a movement driven by technological advances that accelerated data generation to a pace sufficiently fast to move it beyond the capacity of existing resources centralized in a single company or institution. Although this accelerated pace of growth raises many computational problems, it also introduces potential benefits, as innovations induced by big data problems will certainly lead to a range of entirely new scientific discoveries that would not be possible via conventional approaches.

In chemistry, big data can be exploited in many ways; in computer simulations, miniaturized sensors, combinatorial synthesis, in the design of experimental procedures and protocols with increased complexity, in the immediate sharing of experimental results via databases and the Internet, to name but a few. In quantum chemistry, for example, there exists the ioChem-BD platform [9], a tool to manage large volumes of simulation results on chemical structures, bond energies, spin angular momentum, and other descriptive measures. These data volumes demand inspection tools, including versatile browsing and visualization for a minimum level of comprehension, in addition to techniques and tools for systematic analysis. In data-driven medicinal chemistry [2] investigators must face critical issues and factors that scale with the data, such as data sharing, modelling molecular behaviour, implementation and validation with experimental rigor, and defining and identifying ethical considerations.

Big data has been often characterized in terms of the so-called five V's: Volume, Velocity, Variety, Veracity, and Value [1,2]. Albeit not a strong definition, this description somehow captures the characteristic properties of the big data scenario. As far as **volume** is concerned, size is relative across research fields – what is considered big in chemistry may be small in computer science; what really matters is to which extent the data is manageable and usable by those who need to learn from it. Chemistry produces big data volumes by means of techniques such as parallel synthesis [10], high throughput screening (HTS) [11], and first-principle calculations as reported in notorious efforts on quantum chemistry [147], molecules in general as in the AFLOW project [148], and organic molecules as in the ANI-1 project [149]. Big data volumes in chemistry also originate from compilations of the literature and patent repositories [12,13]. Closely related is **velocity**, which refers to the pace of data generation and may affect the capability of drawing conclusions and identifying alternative experimental directions, demanding off-the-shelf analytical tools to support timely summarization and hypotheses validation. **Variety** refers to the diversity of data types and formats currently available. While chemistry has the advantage of an established universal language to describe compounds and reactions, many problems arise when translating this language into computational models whose usage varies across research groups, and even across individual researchers. **Veracity** in chemistry is closely concerned with the potential lack of quality in data produced by imprecise simulations or collected from experiments not conforming to a sufficiently rigid protocol, especially when biological organisms are involved. Finally, **value** refers to the obvious urge for data that is trustworthy, precise and conclusive.

The importance of big data for chemistry is highlighted in several initiatives discussed in Section 3.1, such as the BIGCHEM project described by Tetko et al. [14]. Their work is illustrative of the issues in handling big data in chemistry and life sciences: it includes a discussion on the importance of data quality, the challenges in visualizing millions of data instances and the use of data mining and ML for predictions in pharmacology. Of particular relevance is the search for suitable strategies to explore billions of molecules, which can be useful in various applications, especially in the pharmaceutical industry, in order, for example, to reduce the massive cost of identifying new lead compounds [14].

## 2.2 Machine learning (ML)

In computer science, the standard approach is to use programming languages to code algorithms that "teach" the computer to perform a particular task. ML, in turn, refers to implementing algorithms that tell the computer how to "learn", given a set of data instances (or examples) and some underlying assumptions. Computer programs can then execute tasks that are not explicitly defined in the code. As the very name implies, it depends on learning, a process that in humans takes years, even decades, and that often happens based on the observation of both successes and failures. It is thus implicit that such learning depends on a large degree of experimental support. In its most usual approach, ML depends on an extensive set of successful and unsuccessful examples that will mold the underlying learning algorithm. This is where ML and big data collide. The abundance of both data and computing capacity has brought feasibility to approaches that would not work otherwise due to a lack of sufficient examples to learn from, and/or processing power to drive the learning process. In fact, specialists argue that data collection and preparation in ML can demand more effort than the actual design of the learning algorithms [15]. Nevertheless, solving the issues to build effective learning programs is worthwhile; computers handle datasets much larger than humans can possibly do, as they are not susceptible to fatigue; and, unless mistakenly programmed, they hardly ever make numerical errors.

ML is a useful approach to problems for which designing explicit algorithms is difficult or infeasible, as in the case of spam filtering or detecting meaningful elements in images. Such problems have a huge space of possible solutions; and thus, rather than searching for an explicit solution, a more effective strategy is to have the computer progressively learn new patterns directly from examples. Many chemistry problems conform to this strategy, including protein structure prediction, virtual screening of host-guest binding behaviour, materials design, properties prediction, and the derivation of models for quantitative structure-activity relationship (QSAR). This last one is, itself, an ML practice based on classification and regression techniques; it is used, for example, to predict the biological activity of a compound having its physicochemical properties as input.

### 2.3 An overview of Deep Learning

The latest achievements in ML are related to techniques broadly known as Deep Learning (DL) achieved with Deep Neural Networks (DNN) [17], which are outperforming state-of-the-art algorithms in handling problems such as image and speech recognition (see Figure 1(a)). Nearly all DL algorithms rely on artificial neural networks (ANNs), a biology-inspired technique in which the underlying principle is to approximate complex functions by translating a large number of inputs into a proper output. The principles behind DL are not new, dating back to the introduction of the Perceptron neural network in 1958. After decades of disappointing results in the 1980's and 1990's, ANNs were revitalised with impressive innovations in 2012 in the seminal paper by Krizhevsky et al. [18] and their AlexNet architecture for image classification, inspired by the ideas of LeCun et al. [19]. The driving factors responsible for this drastic change in the profile of a 50-year research field were significant algorithmic advances coupled with huge processing power (thanks to GPU advances), big data sets, and robust development frameworks.

Figure 1 illustrates the evolving popularity of DL and its applications. Figure 1(a) shows the rate of improvement in the task of image classification after the introduction of DL methods. Figure 1(b) shows the increasing interest in the topic, while the increasing popularity of the major software packages [20], Torch, Theano, Caffe, Tensorflow, and Keras, is demonstrated in Figure 1(c) [21].

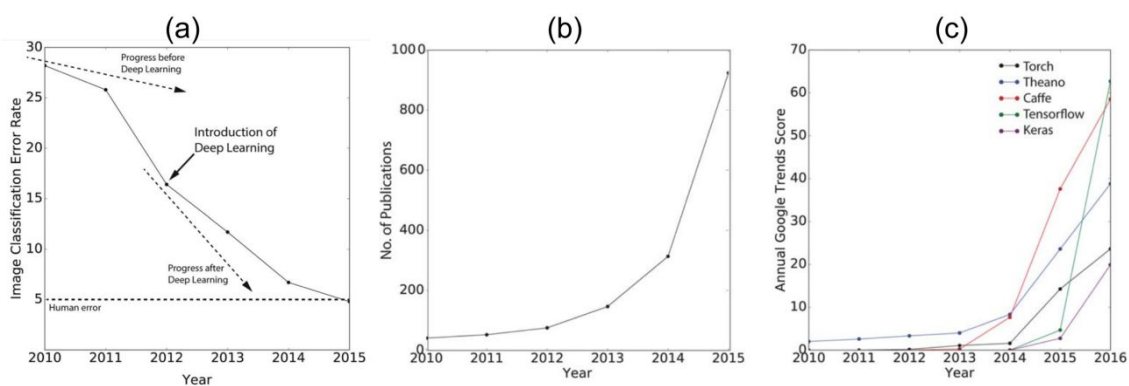

*Figure 1 – Facts about Deep Learning. (a) The improvement of the image classification rate after the introduction of the AlexNet deep neural network. (b) The increasing number of publications as indexed by ISI (International Scientific Indexing). (c) The popularity (Google Trends Score) of major Deep Learning software packages in the current decade: Torch, Theano, Caffe, Tensorflow, and Keras* [20]. *Images reproduced from* [21].

A common method of doing DL is by means of deep feedforward networks, or, simply, feedforward neural networks, a kind of multiplayer perceptron [150]. They work by approximating a function *f\**, as in the case of a classifier *y = f\** that maps an input vector ***x*** to a category *y*. Such mapping is formally defined as *y = f(**x**;**θ**)*, whereas the network must learn the parameters ***θ*** that result in the best function approximation. One can think of the network as a pipeline of interconnected layers of basic processing units, the so-called neurons, which work in parallel; each neuron being a vector-to-scalar function. The model is inspired by Neuroscience findings according to which a neuron receives input from many other neurons; each input is multiplied by a weight – the set of all the weights corresponds to the set of parameters ***θ***. After receiving the vector of inputs, the neuron computes its activation value, a process that proceeds up to the output layer.

Initially, the network does not know the correct weights. To determine the weights, it uses a set of labelled examples, so that every time the classifier *y = f\** misses the correct class, the weights are adjusted by back propagating the error. During this back-propagation, a widely used method to adjust the weights is named Gradient Descent, which calculates the derivative of a loss function (e.g., mean squared error) with respect to each weight and subtract a learning factor from that weight. The adjustment repeats for multiple labelled examples, and over multiple iterations until approximating the desired function. All this process is called *training phase*, and the abundance of labelled data produced nowadays has drastically changed the domains in which ML can work upon.

With an appropriately designed and complex architecture, possibly consisting of dozens of layers and hundreds of neurons, an extensively trained artificial neural network defines a mathematical process whose dynamics is capable of embodying increasingly complex hierarchical concepts. As a result, the networks allow machines to mimic abilities once considered to be exclusive to humans, such as translating text or recognizing objects

Once the algorithms are trained, they should be able to generalise and provide correct answers for new examples of similar nature. There is an interesting optimization issue in balancing the fit to the training data. Overfitting and the model will make correct mappings only to the training examples; underfitting and the model will miss even previously seen examples. For sensor networks, for example, overfitting produces very low error with respect to the training set; it happens because the model encompasses both the noise and the real signal. This often results in a system that generalizes poorly, as the noise is random. So a compromise is required that involves using several different training sets [22] and, most important, regularization techniques [150].

### 2.4 The flow path towards data-based scientific discovery in chemistry

Figure 2 illustrates the standard flow path from data production to the outcome of ML. The first step concerns methods to produce sufficient data to feed computational learning methods. Such data must initially be analysed by a domain expert, who will classify, label, validate, or reject the results of an experiment or simulation. The pre-processing step can be quite laborious and critical in that, if not taken rigorously, it might invalidate the remainder of the process or compromise its results. Once the data is ready, it is necessary to iterate it over an ML method; such methods require a stage of learning (or training) in which the computer learns from the known results provided by the domain expert. During the training step, knowledge is transferred from the training data to the computer in the form of algorithmic settings that are specific to each method. A model is then learned, i.e., a mathematical abstraction that, if computationally executed, can be applied to new, unseen, data, and produce accurate classifications or regressions as a final outcome of the process.

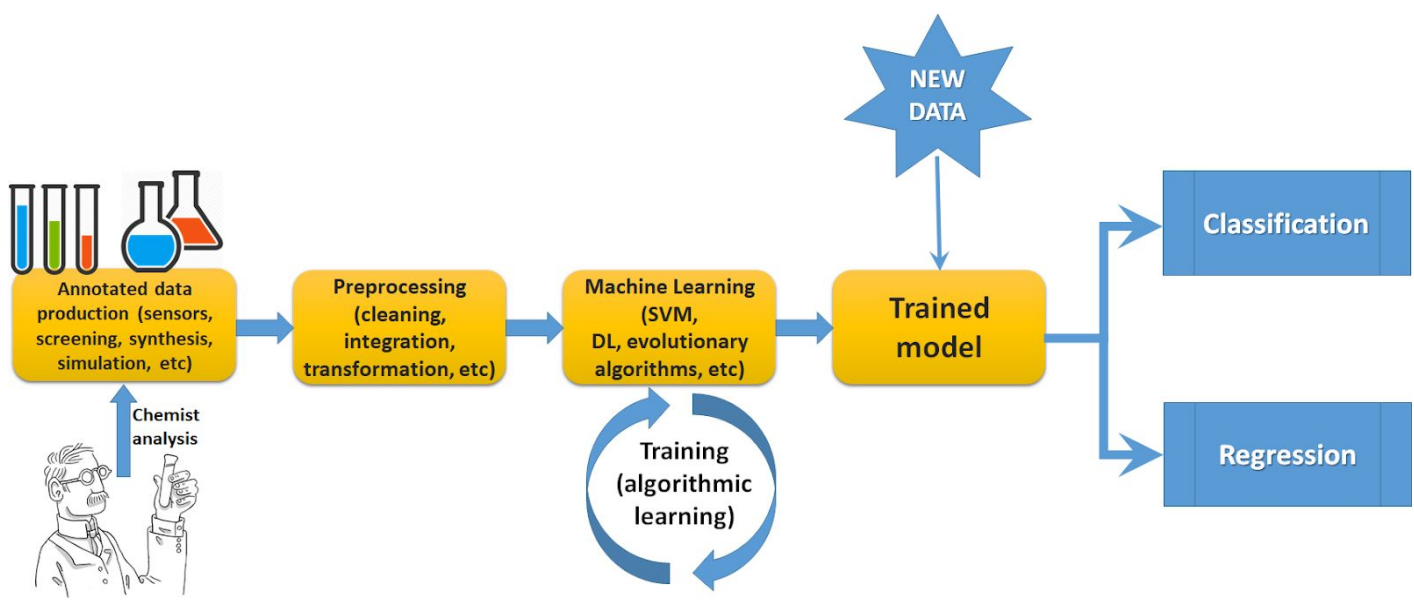

*Figure 2 – The standard flow path of ML, from data production to classification or inference.*

## 3. Materials Discovery

Using computational tools to discover new materials and evaluate material properties is an old endeavour. For instance, DENDRAL, the first documented project on computer-assisted organic synthesis, dates back to 1965 [23]. According to Szymkuc et al. [24] the early enthusiasm observed in the initial attempts waned with successive failures to obtain a reasonable predictive power that could be used to plan organic synthesis; so much so that teaching a computer to perform this task was at some point considered a "mission impossible" [24]. However, renewed interest has emerged with the enhanced capability afforded by big data and novel ML approaches, particularly since one may envisage – for the first time ever – the possibility of exploring a considerable amount of the space of possible solutions defined by the elements in the periodic table and the laws of reactivity. In doing so, the number of possible material structures is estimated as $10^{100}$, larger than the number of particles in the universe [25]. In this vastness, discovery of new materials must face resource and time constraints. Expensive experiments must be well-planned, ideally targeting lead structures with high potential of generating new materials with useful properties. The power of ML in chemistry offers the greatest potential in this scenario. Of course, this comes at a cost; these complex spaces must be mathematically modelled, and a significant number of the representative patterns must be available as examples for the ML system to learn from.

Predicting materials properties from basic physicochemical properties involves exploring quantitative structure-property relationship (QSPR), the analogous version of QSAR for non-biological applications [21,26,27]. There are many examples of ML applied in this domain. The reaction outcomes from the crystallization of templated vanadium selenites were predicted with a *Support Vector Machine* (SVM)-based model where the training set included a "dark" portion of unsuccessful reactions compiled from laboratory notebooks [28] – shortly, SVM refers to a discriminative classifier formally defined by a separating hyperplane, a widely used technique in ML.

Prediction of the target compound was obtained with a success rate of 89%, higher than that obtained with human intuition (78%) [28]. For inorganic solid-state materials, atom-scale calculations have been a major tool to help understand materials behaviour and accelerate materials discovery [29]. Some of the relevant properties, however, are only obtained at a very high computational cost, which has stimulated the use of data-based discovery. In addition to highlighting major recent advances, Ward and Wolverton [29] comment upon current limitations in the field, such as the limited availability of appropriate software targeted at the computation of materials properties.

In the choice of an outline to describe contributions found in the chemistry literature related to materials discovery, we selected a subset of topics that exemplify the many uses of ML.

### 3.1 Large Databases and Initiatives

The materials genome initiatives, multi-institutional international efforts seeking to establish a generic platform for collaboration [30,31], set a hallmark for the importance of big data and ML in chemistry and materials sciences. A major goal is perhaps to move beyond the trial-and-error empirical approaches prevailing in the past [21]. For example, the US Materials Genome Initiative (MGI) (https://www.mgi.gov/) established the following issues as major challenges [32]:

• *Lead a culture shift in materials research to encourage and facilitate an integrated team approach;*
• *Integrate experiment, computation, and theory and equip the materials community with advanced tools and techniques;*
• *Make digital data accessible; and*
• *Create a world-class materials workforce that is trained for careers in academia or industry.*

To encourage a cultural change in materials research, ongoing efforts intend to generate data that can serve both to validate existing models and to create new, more sophisticated models with enhanced predictive capabilities. This has been achieved with a virtual high-throughput experimentation facility involving a national network of labs for synthesis and characterization [32], and partnerships between academia and industry for tackling specific applications. For example, the *Center for Hierarchical Materials Design* is developing databases for materials properties and materials simulation software [32], and alliances have been established to tackle topics like *Materials in Extreme Dynamic Environments,* and *Multi-Scale Modelling of Electronic Materials* [32]. There are also studies of composite materials to improve aircraft fuel efficiency and of metal processing to produce lighter weight products and vehicles [32]. Regarding the integration of experiment, theory and computer simulation, perhaps the most illustrative example is an automated system designed to create a material, test it and evaluate the results, after which the best next experiment is chosen in an iterative procedure. The whole process is conducted without human intervention. The system is already in use to speed up the development process of high-performance carbon nanotubes for use in aircraft [32]. Another large-scale program at the University of California, Berkeley, used high-performance computing and state-of-the-art theoretical tools to produce a publicly-available database of the properties of 66,000 new and predicted crystalline compounds and 500,000 nanoporous materials [32].

There are cases that require combining different levels of theory and modelling with experimental results, particularly for more complex materials. In the *Nanoporous Materials Genome Center,* microporous and mesoporous metal-organic frameworks and zeolites are studied for energy-relevant processes, catalysis, carbon capture, gas storage, and gas- and solution-phase separation. The theoretical and computational approaches range from electronic structure calculations combined with Monte Carlo sampling methods, to graph theoretical analysis, which are assembled into a hierarchical screening workflow [32].

An important feature of MGI is the provision of infrastructure for researchers to report their data in a way that it can be curated. Programs such as *Materials Data Repository* (MDR) and *Materials Resource Registry* are being developed to allow for worldwide discovery, in some cases based on successful resources from other communities, such as the registry of the Virtual Astronomical Observatory [32]. This component of the "Make digital data accessible" goal of MGI has already provided extensive datasets, e.g. for compounds (~1,500 compounds) to be used in electrodes for ion-lithium batteries and over 21,000 organic molecules for liquid electrolytes. These programs ensure the immediate access by the industry to data that may help to accelerate materials development in applications such as hydrogen fuel cells, pulp, and paper industry and solid-state lighting [32]. Also worth mentioning is the QM database containing the ground-state electronic structures for 3 million molecules and 10 low-lying excited states for more than 3.5 million molecules (e.g., "water", "ethanol", "ethyl alcohol", etc) in the "PubChemQC" project (http://pubchemqc.riken.jp/) [33]. For this database, the ground state structures were calculated with density-functional theory (DFT) at the B3LYP/6-31G* level, while the time-dependent DFT with B3LYP functional and 6-31+G* basis set was used for the excited states. The project also employs ML (SVMs and regression) for predicting DFT results related to the electronic structure of molecules.

### 3.2 Identification of compounds with genetic algorithms

In bio-inspired computation, computer scientists define procedures that mimic mechanisms observed in natural settings. This is the case of the genetic algorithms [151] inspired on Charles Darwin's ideas of evolution, which mimic the "survival of the fittest" principle to set up an optimization procedure. In these algorithms, known functional compounds are crossed over along with a mutation factor to produce novel compounds; the mutation factor introduces new properties into the mutations. Novel compounds with no useful properties are disregarded, while those displaying useful properties (high fitness) are selected to produce new combinations. After a certain number of generations (or iterations), new functional compounds emerge with some properties inherited from their ancestors, supplemented with other properties acquired along their mutation pathway. Of course, this is an oversimplified description of the process, which depends on accurate modelling of the compounds, a proper definition of the mutation procedure, and a robust evaluation of the fitness property. The latter may arrive by means of calculations, as in the case of conductivity or hardness, reducing the need for expensive experimentation.

In genetic algorithms, each compositional or structural characteristic of a compound is interpreted as a gene. Examples of chemical genes include fraction of individual components in a given material, polymer block sizes, monomer compositions, and processing temperature. The genome refers to the set of all the genes in a compound, while the resulting properties of a genome are named a phenotype. The task of a genetic algorithm is to scan the search space of the genes' domains to identify the most suitable phenotypes, as measured by a fitness function. The relationship between the genome and the phenotype gives rise to the fitness landscape (see Tibbetts et al. [34] for a detailed background). Figure 3 illustrates a fitness landscape for two hypothetical genes, say, block size and processing temperature of a polymer synthesis process whose aim is to achieve high rates of hardness. Note that when exploring a search space, the fitness landscape is not known in advance, and rarely only bi-dimensional; instead, it is implicit in the problem model defined by the genes' domains and in the definition of the fitness function. The modelling of the problem is correct if the genes permit a gainful movement over the fitness landscape while the fitness function correlates with interesting physical properties. In the example in Figure 3, the genetic algorithm moves along the landscape by producing new compounds while avoiding compounds that will not improve the fitness. The mechanism of the genetic algorithm, therefore, grants it a higher probability of moving towards phenotypes with the desired properties. For a comprehensive review focused on materials science, please refer to the work of Paszkowicz [152].

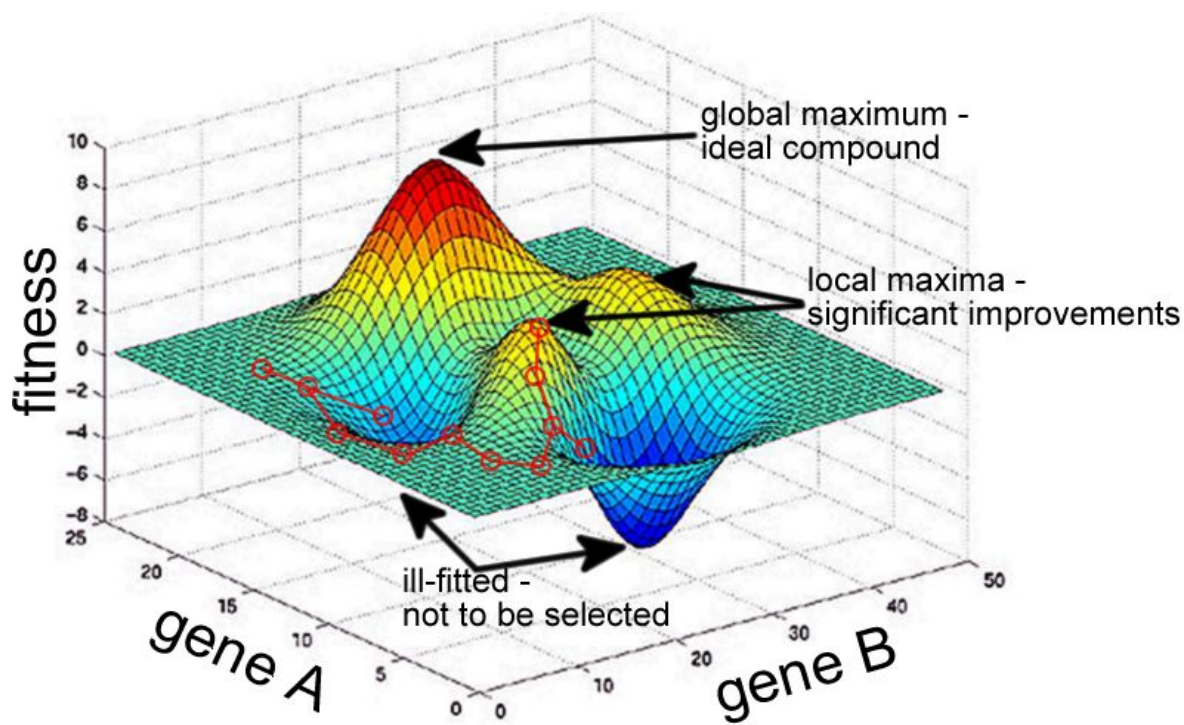

*Figure 3 - A fitness landscape considering two hypothetical genes as, for example, block size and processing temperature of a polymer synthesis. Fitness could be hardness, for instance. The landscape contains local maxima, a global maximum, and a global minimum. In red, the path of a genetic algorithm along 11 iterations.*

Identification of more effective catalysts has also benefited from genetic algorithms, as in the work by Wolf et al. [35] where a set of oxides ($B_2O_3$, $Fe_2O_3$, $GaO$, $La_2O_3$, $MgO$, $MnO_2$, $MoO_3$, and $V_2O_5$) was taken as the initial population of an evolutionary process. Aimed at finding the catalysts that would optimize the conversion of propane into propene through dehydrogenation, the elements of the initial set were iteratively combined to produce four generations of catalysts. In total, the experiment produced 224 new catalysts with an increase of 9% in conversion (T=500ºC, $C_3H_8/O_2$, $p(C_3H_8)$=30Pa). A thorough review of evolutionary methods in searching for more efficient catalysts was given by Le et al. [25]. Bulut et al. [36] explored the use of polyimide solvent-resistant nanofiltration membranes (phase inversion) to produce membrane-like materials. The aim was to optimize the composition space given by two volatile solvents (tetrahydrofuran and dichloromethane) and four non-solvent additives (water, 2-propanol, acetone, and 1-hexanol). This system was modelled as a genome with eight variables corresponding to a search space of $9 \times 10^{21}$ possible combinations, which could not be exhaustively scanned no matter the screening method. The solution was to employ a genetic algorithm driven by a fitness function defined by the membrane retention and permeance. Throughout four generations and 192 polymeric solutions, the fitness function indicated an asymptotic increase in the membrane performance.

### 3.3 Synthesis prediction using ML

The synthesis of new compounds is a challenging task, especially in organic chemistry. The search for machine-based methods to predict which molecules will be produced from a given set of reactants and reagents started in 1969, with Corey and Wipke [37], who demonstrated that synthesis (and retrosynthesis) predictions could be carried out by a computing machine. Their approach was based on templates produced by expert chemists that defined how atom connectivity would rearrange, given a set of conditions – see Figure 4. Despite demonstrating the concept, their approach suffered from the limited template sets, which prevented their method from encompassing a wide range of conditions, and that would fail in face of even the smallest alterations.

The use of templates (or rules) to transfer knowledge from human experts to computers, as seen in the work of Corey and Wipke, corresponds to an old computer science paradigm broadly referred to as "expert systems" [39]. This approach has attained limited success in the past, due to the burden of producing sufficiently comprehensive sets of rules capable of yielding results over a broad range of conditions, coupled with the difficulty of anticipating exceptional situations. Nevertheless, it has gained renewed interest recently, as ML methods may contribute to automatic rule generation taking advantage of big datasets, as it is being explored, for instance, in medicine [40].

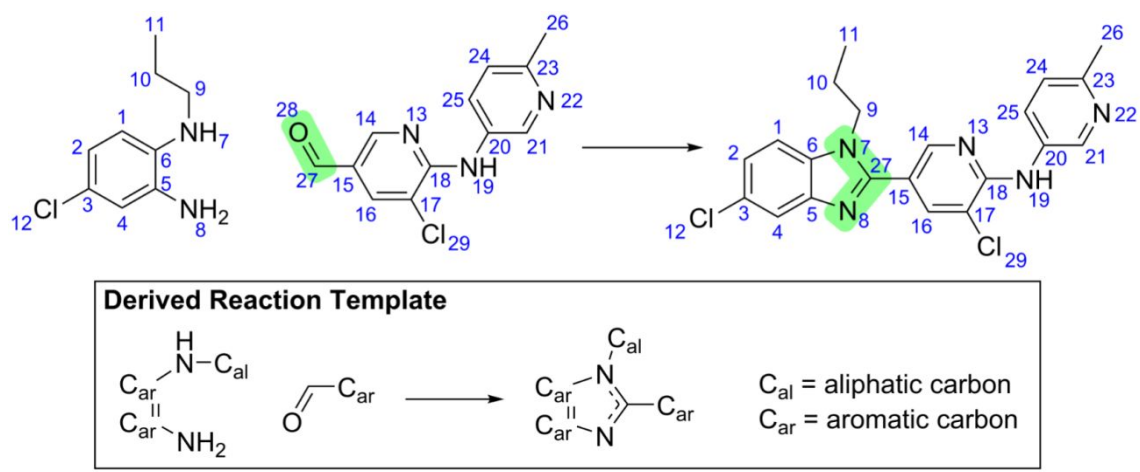

*Figure 4 – Example of a reaction and its corresponding reaction template. The reaction is centered in the green-highlighted areas (27,28), (7,27), and (8,27). The corresponding template includes the reaction center and nearby functional groups. Reproduced from [38].*

In association with big data, ML became an alternative to extracting knowledge not only from experts, but also from datasets. Coley et al. [38], for example, used a 15,000-patents dataset to train a neural network to identify the sequence of templates that would most probably produce a given organic compound during retrosynthesis. Segler and Waller [41] used 8.2 million binary reactions (including 14.4 million molecules) acquired from the Reaxys web-based chemistry database (https://www.reaxys.com) to build a knowledge graph, a bipartite directed graph *G=(M,R,E)*, made of two sets of nodes, where *M* stands for the set of molecules and *R* for the set of reactions, plus one set *E* of labelled edges, each one representing a role *t* ∈ *{reactant, reagent, catalyst, solvent, product}*. A schema of the approach is depicted in Figure 5. A *link prediction* ML algorithm was employed, which predicts new edges from the characteristics of existing paths within the graph structure given by edges of type *reactant*. In the example depicted in Figure 5(b), the reactant path between molecules 1 and 4 indicates a missing reaction node between them, i.e. node D in Figure 5(c). The experiments confirmed a high accuracy in predicting the products of binary reactions, and in detecting reactions that are unlikely to occur.

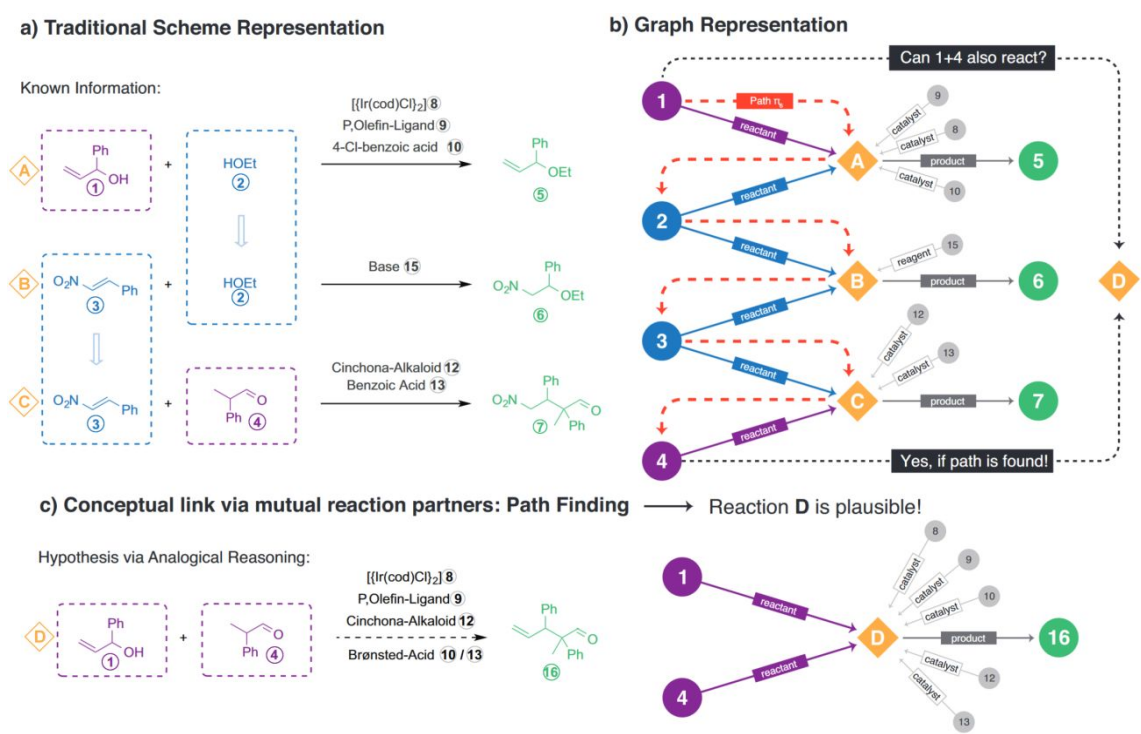

*Figure 5 – Graph representation of reactions. (a) Four molecules, 1, 2, 3, and 4 and three reactions, A, B, and C. (b) The graph representation in which the circles represent the molecule nodes, and the diamonds represent the reaction nodes; edges describe the role, reactant, reagent, catalyst, solvent, or product, of each molecule to a given reaction. Notice the path, depicted in orange color, made of reactant edges between molecules 1 and 4. (c) The missing reaction node D that was indicated by the reactant path found between 1 and 4. Reproduced from* [41].

Owing to the limitations of *ad hoc* procedures relying on templates, Szymkuc et al. [24] advocated that for chemical syntheses the chemical rules from stereo- and regiochemistry may be coded with elements of quantum mechanics to allow ML methods to explore pathways of known reactions from a large database. This obviously tends to increase the data space to be searched. From recent literature, not restricted to the chemical field, DL appears as the most promising approach for successfully exploring large search spaces. Schwaller et al. [42] used DL methods to predict outcomes of chemical reactions and found the approach suitable to assimilate latent patterns for generalizing out of a pool of examples, even though no explicit rules are produced. Assuming that organic chemistry reactions sustain properties similar to those studied in linguistic theories [43], they explored state-of-the-art neural-networks to translate reactants into products similarly to how translation is performed from one language into another. In their work, a DNN was trained over the Lowe's dataset of US patents, which contains patents applied between 1976 and 2016 [44], including 1,808,938 reactions described using the SMILES [45] chemical language, which defines a notation system to represent molecular structures as graphs and strings amenable to computational processing. The Jin's dataset [46], a cleaned version of the Lowe's dataset after removing duplicates and erroneous reactions, with 479,035 reactions, was also used. They achieved an accuracy of 65.4% for single-product reactions over the entire Lowe's dataset, and an accuracy of 80.3% over the Jin's dataset. Bombarelli et al. [47] also combined the SMILES notation and a DNN to map discrete molecules into a continuous multidimensional space in which the molecules are represented as vectors. In such a continuous space, it is possible to predict properties of the existing vectors and predict new vectors with certain properties. Gradients are computed to indicate where to look for vectors whose properties vary in a desired way, and optimization techniques can be employed to search the best candidate molecules. After finding new vectors, a second neural network converts the continuous vector representation into a SMILES string that reveals a potential lead compound. DNNs were also employed to predict reactions with 97% accuracy on a validation set of 1 million reactions, a clearly superior performance to previous rule-based expert systems [48].

### 3.4 Quantum chemistry

The high computational cost of quantum chemistry has been a limiting factor to exploring the virtual space of all the possible molecules from a quantum perspective. This is the reason why researchers are increasingly resorting to ML approaches [49]. Indeed, ML has been used to replace or supplement quantum mechanical calculations for predicting parameters such as the input for semi-empirical QM calculations [50], modelling electronic quantum transport [51], or establishing a correlation between molecular entropy and electron correlation [49]. ML can be employed to overcome or minimize the limitations of *ab initio* methods such as DFT, which are useful to determine chemical reactions, quantum interactions between atoms, molecular and materials properties, but are not suitable to treat large or complex systems [52].

The coupling of ANNs and *ab initio* methods is exemplified in the PROPhet project [52] (short for PROPerty Prophet), for establishing non-linear mappings between a set of virtually any system property (including scalar, vector, and/or grid-based quantities) to any other one. PROPhet provides, among its functionalities, the ability to learn analytical potentials, non-linear density functions, and other structure-property or property-property relationships, reducing the computational cost of determining materials properties, in addition to assisting design and optimization of materials [52]. Quantum-chemistry-supported ML approaches have also been used to predict the sites of metabolism for the cytochrome P450 with a descriptor scheme where a potential reaction site was identified by determining the steric and electronic environment of an atom and its location in the molecular structure [53].

ML algorithms can accelerate the determination of molecular structures via DFT, as described by Pereira et al. [54], who estimated HOMO and LUMO orbital energies using molecular descriptors based only on connectivity. Another aim was to develop new molecular descriptors, for which a database containing >111,000 structures was employed in connection with various ML models. With random forest models, the mean absolute error (MAE) was smaller than 0.15 and 0.16 eV for HOMO and LUMO orbitals, respectively [54]. The quality of estimations was considerably improved if the orbital energy calculated by the semi-empirical PM7 method was included as an additional descriptor.

The prediction of crystal structures is among the most important applications of high-throughput experiments, which rely on *ab initio* calculations. DFT has been combined with ML to exploit interatomic potentials for searching and predicting carbon allotropes [55]. In this latter method, the input structural information comes from liquid and amorphous carbon only, with no prior information of crystalline phases. The method can be associated with any algorithm for structure prediction, and the results obtained using ANNs were orders of magnitude faster than with DFT [55].

With a high-throughput strategy, time-dependent DFT was employed to predict the electronic spectra of 20,000 small organic molecules, but the quality of these predictions was poor [56]. Significant improvement was attained with a specific ML method named Ansatz, which was employed to determine low-lying singlet-singlet vertical electronic spectra, with excitation reproduced to within ± 0.1 eV for a training set of 10,000 molecules. Significantly, prediction error decreased monotonically with the size of the training set [56]; this experiment opened the prospect for addressing the considerably more difficult problem of determining transition intensities. As a proof-of-principle exercise, accurate potential energy surfaces and vibrational levels for methyl chloride were obtained in which *ab initio* energies were required for some nuclear configurations in a predefined grid [57]. ML using a self-correcting approach based on kernel ridge regression was employed to obtain the remaining energies, reducing the computational cost of the rovibrational spectra calculation by up to 90%, since tens of thousands of nuclear configurations could be determined within seconds [57]. ANNs were trained to determine spin-state bond lengths and ordering in transition metal complexes, starting with descriptors obtained with empirical inputs for the relevant parameters [58]. Spin-state splittings of single-site transition metal complexes could be obtained to within 3 kcal mol$^{-1}$; an accuracy comparable to that of DFT calculations. In addition to predicting structures validated with *ab initio* calculations, the approach is promising for screening transition metal complexes with properties tailored for specific applications.

The performance of ML methods for given applications in quantum chemistry has been assessed via contests, as often done in computer science. An example is the Critical Assessment of Small Molecule Identification (CASMI) Contest (http://ww.casmi-contest.or) [59], in which ML and chemistry-based approaches were found to be complementary. Improvements in fragmentation methods to identify small molecules were considerable, and this should further improve in the coming years with the integration of further high-quality experimental training data [59].

According to Goh et al.[21], DNNs have been used in quantum chemistry so far to a more limited extent than they have in computational structural biology and computer-aided drug design, possibly because the extensive amounts of training data they require may not yet be available. Nevertheless, Goh et al. state that such methods will eventually be applied massively for quantum chemistry, owing to their observed superiority in comparison to traditional ML approaches – an opinion we entirely support. For example, DNNs applied to massive amounts of data could be combined with QM approaches to yield accurate QM results for a considerably larger number of compounds than is feasible today [49].

Though the use of DNN in quantum chemistry may be still at an embryonic stage, it is possible to identify significant contributions. Tests have been made mainly with the calculation of atomization energies and other properties of organic molecules [21] using a portion of 7,000 compounds from a library of $10^9$ compounds, where the energies in the training set were obtained with the PBE0 (Perdew–Burke-Ernzerhof (PBE) exchange energy and Hartree-Fock exchange energy) hybrid function [60]. DNN models yielded superior performance compared to other ML approaches, since a DNN could successfully predict static polarizabilities, ionization potentials and electron affinity, in addition to atomization energies of the organic molecules [61]. It is significant that the accuracy was similar to the error of the corresponding theory employed in QM calculations to obtain the training set. Applying DNNs to the dataset of the Harvard Clean Energy Project to discover organic photovoltaic materials, Aspuru-Guzik et al. [62] predicted HOMO and LUMO energies and power conversion efficiency for 200,000 compounds, with errors below 0.15 eV for the HOMO and LUMO energies. DL methods have also been exploited in predicting ground- and excited state properties for thousands of organic molecules where the accuracy for small molecules can be even superior to QM *ab initio* methods [61].

### 3.5 Computer-aided Drug Design

Drug design has relied heavily on computational methods in a number of ways, from computer calculations of quantum chemistry properties with *ab initio* approaches, as discussed in the previous section, to screening processes in high-throughput analysis of families of potential drug candidates. Huge amounts of data have been gathered over the last few decades, with a range of experimental techniques, which may contain additional information on the materials' properties. This is the case of mass spectrometry datasets that may contain valuable hidden information on antibiotics and other drugs. In the October 2016 issue of *Nature Chemical Biology* [63], the use of big data concepts was highlighted in the discovery of bioactive peptidic natural products via a method referred to as DEREPLICATOR. This tool works with statistical analysis via Markov Chain-Monte Carlo to evaluate the match between spectra in the database of Global Natural Products Social infrastructure (containing over one hundred million mass spectra) with those from known antibiotics. Crucial for the design of new drugs are their absorption, distribution, metabolism, excretion, and toxicity (ADMET) properties on which the pharmacokinetic profile depends [64]. Determining ADMET properties is not feasible when such a large number of drug candidates are to be screened; therefore, computational approaches are the only viable option. For example, prediction results generated from various QSAR models can be compared to experimentally-measured ADMET properties from databases [65–71]. These models are limited in that they may not be suitable to explore novel drugs, which motivates an increasing interest in ML methods, which can be trained to generate predictive models that may discover implicit patterns from new data used to determine more accurate models [64].

Pires and Blundell developed the approach named pkCSM (http://structure.bioc.cam.ac.uk/pkcsm), in which ADMET properties of new drugs can be predicted with graph-based structural signatures [64]. In pkCSM the graphs are constructed by representing atoms as nodes while the edges are given by their covalent bonds. Also essential are the labels used to decorate the nodes and edges with physicochemical properties, similarly to the approach used in embedded networks [3]. The concept of structural signatures is associated with establishing a signature vector that represents the distance patterns extracted from the graphs [64]. The workflow for pkCSM is depicted in Figure 6, which involves two sets of descriptors for input molecules: general molecule properties and the distance-based graph signature. Molecular properties used include lipophilicity, molecular weight, surface area, toxicophore fingerprint and the number of rotatable bonds.

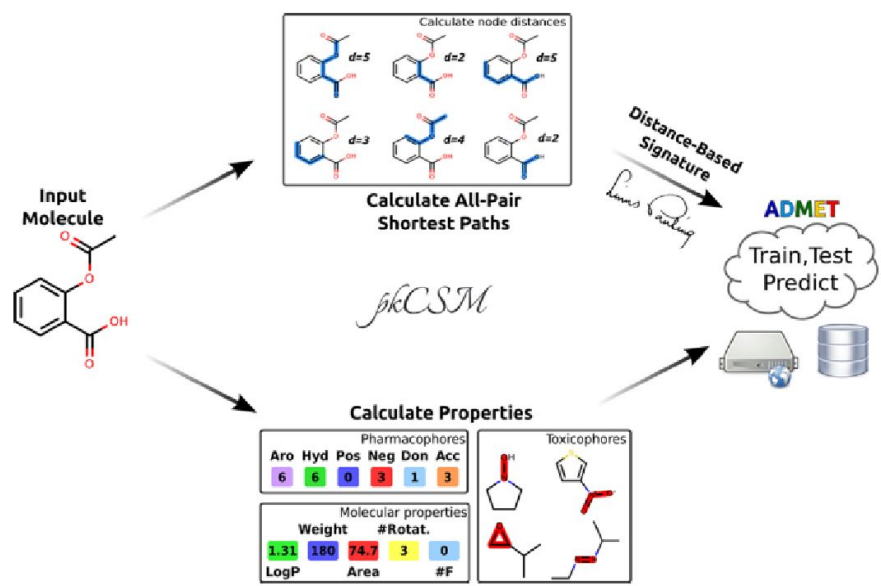

*Figure 6 – The workflow of pkCSM is represented by the two main sources of information, namely the calculated molecular properties and shortest paths, for an input molecule. With these pieces of information, the ML system is trained to predict ADMET properties. Reproduced from [64].*

QSAR [16,72–74] is ubiquitous in some of these applications, as in the computer-aided drug design that predicts the biological activity of a molecule. The inputs typically are the physicochemical properties of the molecule. The use of DL for QSAR is relatively recent, as typified in the Merck challenge [75], wherein the activity of 15 drug targets was predicted in comparison to a test set. In later work, this QSAR experiment was repeated with a dataset curated from PubChem containing over 100,000 data points, for which 3,764 molecular descriptors per molecule were used as DNN input features [76]. DNN models applied to the Tox21 challenge provided the highest performance [77], with 99% of neurons in the first hidden layer displaying significant association with toxicophore features. Therefore, DNNs may possibly be used to discover chemical knowledge by inspection of the hidden layers [77]. Virtual screening is also relevant to complement docking methods for drug design, as exemplified with DNNs to predict the activity of molecules in protein binding pockets [78]. In another example, Xu et al. [79] employed a dataset with 475 drug descriptions to train a DNN to predict whether a given drug may induce liver injury. They used their trained predictor over a second dataset with 198 drugs of which 172 were accurately classified with respect to their liver toxicity. This type of predictor affords significant time and cost savings by rendering many experiments unnecessary. This practice in the QSAR domain is potentially useful for the non-biological quantitative structure-property relationship (QSPR) [27], in which the goal is to predict physical properties departing from simpler physicochemical properties. Despite the existence of interesting works on the topic [58,62], there is room of for further research.

A highly relevant issue that can strongly benefit from novel procedures standing on big data and classical ML methods is drug discovery for neglected diseases. Cheminformatics tools have been assembled into a web-based platform in the project *More Medicines for Tuberculosis* (MM4TB), funded by the European Union [80]. The project relies on classical ML methods (*Bayesian modelling*, SVMs, *Random Forest*, and *bootstrapping*), collaboratively working on data acquired from the screening of natural products and synthetic compounds against the microorganism *Mycobacterium Tuberculosis*.

Self-organizing maps (SOM) are a particular type of ANN that has been proved useful in a rational drug discovery process, where they assist in predicting the activity of bioactive molecules and their binding to macromolecular targets [81]. Antimicrobial peptide (AmPs) activity was predicted using an adaptive neural network model having the amino acid sequence as input data [82]. The algorithm iterated to optimize the network structure, in which the number of neurons in a layer and their connectivity were free variables. High charge density and low aggregation in solution were found to yield higher antimicrobial activity. In another example of antimicrobial activity prediction, ML was employed to determine the activity of 78 sequences of antimicrobial peptides generated through a linguistic model. In this, the model treats the amino-acid sequences of natural AmPs as a formal language described by means of a regular grammar [83]. The system was not efficient in predicting the 38 shuffled sequences of the peptides, a failure attributed to their low specificity. The authors [83] concluded that complementary methods with high specificity are required to improve prediction performance. An overview of the use of ANNs for drug discovery, including DL methods, is given by Gawehn et al. [84].

Though to a lesser extent than for drug design, ML is also being employed for modelling drug-vehicle relationships, which are essential to minimize toxicity [85]. Authors employed ML on data from the *National Institute of Health's* (NIH) *Developmental Therapeutics Program* to build classification models and predict toxicity profiles for a drug. That is to say, they employed the Random Forest classifier to determine which drug carriers led to the least toxicity with a prediction accuracy of 80%. Since this method is generic and may be applied to wider contexts, we see great potential in its use in the near future owing to the increasing possibilities introduced by nanotech-based strategies for drug delivery.

As in many other application areas, beyond chemistry or pharmaceutical research, the performance of distinct ML methods has been evaluated according to their performance in solving a common problem. It now seems that DL may perform better than other ML methods [86], especially in cases where large datasets have been compiled over the years. Ekins [86] listed a number of applications of DNNs in the pharmaceutical field, including prediction of aqueous solubility of drugs, drug-induced liver injury, ADME and target activity, cancer diagnosis. In a more recent work, Korotcov and collaborators [87] have shown that DL yielded superior results compared to SVMs and other ML approaches in the prediction ability for drug discovery and ADME/Tox data sets. Results were presented for the Chagas' disease, tuberculosis, malaria and bubonic plague.

DL has also succeeded in the problem of protein contact prediction [88]. In 2012, Lena et al. [89] superseded the previously impassable mark of 30% accuracy for the problem. They used a recursive neural network trained over a 2,356-element dataset from the ASTRAL database [90], a big data compendium of protein sequences and relationships. Then, they tested their network over 364 protein folds, achieving the first-time-ever mark of 36% accuracy, which brought new hope to this complex field.

## 4. Sensors and Biosensors for Intelligent Systems and Internet of Things

The term Internet of Things (IoT) was coined at the end of the 20$^{th}$ Century to mean that any type of device could be connected to the Internet, thus enabling tasks and services to be executed remotely. In other words, the functioning of a device, appliance, etc. could be monitored and/or controlled via the Internet. If (almost) any object can be connected, three immediate consequences can be identified: i) sensing must be ubiquitous; ii) huge volumes of data will be generated; iii) systems will be required to process the data and make use of the network of connected "things" for specific purposes. There is a virtually endless list of possible services, ranging across traffic control, health monitoring, surveillance, precision agriculture, control of manufacturing processes, weather monitoring, to name just a few. In an example of sensors and sensing networks for monitoring health and environment with wearable electronics, Wang et al. [91] emphasized the need to develop new materials for meeting the stringent requirements to develop IoT-related applications.

A comprehensive review on chemical sensing (or IoT) is certainly beyond the scope of this paper, and we shall, therefore, restrict ourselves to providing some illustrative examples on how chemical sensors are producing big data, to make the point that sensors and biosensors are key to providing the data needed to solve problems by means of ML. Indeed, methods akin to big data and ML have been employed for analysing data from sensors and biosensors, through computer-assisted diagnosis for the medical area, and other areas where diagnosis relies on sensing devices, such as in fault prediction in industrial settings.

### 4.1 ML in Sensor Applications

Chemistry is essential for IoT sensing and biosensing, as well as in intelligent systems, for a variety of reasons, including the development of new materials for building innovative chemical (and electrochemical) sensing technologies (see, for instance, the review paper by Oliveira Jr et al. [92]). In recent decades, increasingly complex chemical sensing has produced data volumes from a wide range of analytical techniques. There has been a tradition in chemistry – probably best represented by contributions in chemometrics – to employ statistics and computational methods to treat not only sensing data, but also other types of analytical data. Electronic noses (e-noses) are one illustrative example of the use of ML methods in sensing and biosensing. Ucar et al. [93] built an android nose to recognize the odor of six fruits by means of sensing units made of metal-oxide semiconductors whose output was classified using the method *Kernel Extreme Learning Machines* (KELMs). In another work [94], the authors introduced a framework for multiple e-noses with cross-domain discriminative subspace learning, a more robust architecture for a wider odor spectrum. Robust e-sensing is also present in the work of Tomazzoli et al. [95], who employed multiple classification techniques, such as *Partial Least Squares-discriminant analysis* (PLS-DA), *k-Nearest Neighbors* (kNN), and *Decision Trees*, to distinguish between 73 samples of propolis collected over different seasons based on the UV-Vis spectra of hydroalcoholic extracts. The relevance of this study lies in establishing standards for the properties of propolis, a biomass produced by bees and widely employed as an antioxidant and antibiotic due to its amino acids, vitamins, and bioflavonoids. As with many natural products, propolis displays immense variability, including dependence on the season when it is collected, so that excellent quality control must be ensured for reliable practical use in medicine. Automated classification approaches represent, perhaps, the only possible way to attain low-cost quality control for natural products that are candidate materials for cosmetics and medicines.

Disease monitoring and control are essential for agriculture, as is the case of orange plantations, particularly when diseases and deficiencies may yield similar visual patterns. Marcassa and co-workers [96] took images obtained from fluorescence spectra and employed SVMs and ANNs to distinguish between samples affected by the Huanglongbing (HLB) disease and those with zinc-deficiency stress. The ability to process large amounts of data and identify patterns allows one to integrate sensing and classifying tasks into portable devices such as smartphones. Mutlu et al. [97], for instance, used colored-strip images corresponding to distinct pH values to train a *Least-Squares*-SVM classifier. The results indicated that the pH values were determined with a high accuracy.

The visualization of bio(sensing) data to gain insight, support decisions or, simply, acquire a deeper understanding of the underlying chemical reactions has been exploited extensively by several research groups, as reviewed in the papers by Paulovich et al. [98] and Oliveira et al. [99]. Previous results achieved by applying data visualization techniques to different types of problems point to potentially valuable traits when chemical data is inspected from a graphical perspective. Possible advantages of this approach include:

i) the whole range of features describing a given dataset of sensing experiments can be used as the input to multidimensional projection techniques, without discarding information at an early stage that might otherwise be relevant for a future classification task. For example, in electrochemical sensors, rather than using information about oxidation/reduction peaks, entire voltammograms may be considered; in impedance spectroscopy, instead of taking the impedance value at a given frequency, the whole spectrum can be processed in obtaining a visualization.

ii) other multidimensional visualization techniques such as Parallel Coordinates [100] allow identification of the features that contribute most significantly to the distinguishing ability of the bio(sensor).

iii) Various multidimensional projection techniques are available, including non-linear models, which in some cases have been proven to be efficient for handling biosensing data [98]. Such usage is exemplified with an example in which impedance-based immunosensors were employed to detect the pancreatic cancer biomarker CA19-9 [101].

The feature selection mechanism used by Thapa et al. [101] performed via manual visual inspection, and combined with the silhouette coefficient (a measure of cluster quality), was demonstrated to enhance the immunosensor performance. However, more sophisticated approaches can be employed, as in the work by Moraes et al. [102], in which a genetic algorithm was applied to inspect the real and imaginary parts of the electric impedance measured by two sensing units. The method was found capable of distinguishing triglycerides and glucose by means of well-characterized visual patterns. Using predictive modelling with decision trees, Aileni [103] introduced a system named VitalMon, designed to identify correlations between parameters from biomedical sensors and health conditions. An important tenet of the design was data fusion from different sources, e.g. a wireless network, and sensed data related to distinct parameters, such as breath, moisture, temperature, and pulse. Within this same approach, data visualization was combined with ML methods [104] for the diagnosis of ovarian cancer, using input data from mass spectroscopy.

This type of analytical approach is key for electronic tongues (e-tongues) and e-noses, as these devices take the form of arrays of sensing units and generate multivariate data [105]. For example, data visualization and feature selection were combined to process data from a microfluidic e-tongue to distinguish between gluten-free and gluten-containing foodstuff [106]. ML methods to can also be employed to teach an e-tongue whether a taste is good or not, according to the human perception. This has been done for the capacitance data of an e-tongue applied to Brazilian coffee samples, as explained by Ferreira et al.[107]. In that paper, the technique yielding highest performance was referred to as an ensemble feature selection process, based on the *Random Subspace Method* (RSM). The suitability of this method for predicting coffee quality scores, from the impedance data obtained with an e-tongue, was supported by the high correlation between the predicted scores and those assigned by a panel of human experts.

**4.2 Providing data for big data and ML applications with chem/bio sensor networks**

As discussed, sensors and biosensors are crucial to provide the information at the core of big data and ML. Large-scale deployments of essentially self-sustaining wireless sensor networks (WSNs) for personal health and environment monitoring, whose data can be mined to offer a comprehensive overview of a person's or ecosystem's status, have been anticipated from long ago [108]. In this vision, large numbers of distributed sensors continuously collect data that is further aggregated, analysed, and correlated to report upon real-time changes in the quality of our environment or an individual's personal health. At present, deployments of chemical WSNs are limited in scale, and most of the sensors employed rely on modulation of physical properties, such as temperature, pressure, conductivity, salinity, light illumination, moisture, or movement/vibration, rather than chemical measurements. In environmental monitoring, there are examples of relatively large-scale deployments which encompass forest surveillance (*e.g.* GreenOrbs WSN with approximately 5,000 sensors connected to same base station [109]), vineyard monitoring [110–113], volcanic activity monitoring [114], monitoring greenhouses [115,116], soil moisture monitoring [117], water status monitoring [118], animal migrations [119,120] and marine environment monitoring [121,122], amongst others [123]. In the wireless body sensor networks (WBSN) arena, with over two-thirds of the world's population already connected by mobile devices [124], the potential impact of WSNs and IoT on human performance, health and lifestyle is enormous. While numerous wearable technologies specific to fitness, physical activity and diet are available, there are studies indicating that devices that monitor and provide feedback on physical activity may not offer any advantage over standard approaches [125]. These studies suggest that perhaps ML approaches will be required to generate a meaningful and effective improvement in an individual's lifestyle. However, physical sensors offer only a limited perspective of the environment status or individual's condition. A much fuller picture requires more specific molecular information, an arena where WSNs based on chemical and biochemical sensors are essential to bring the IoT to the next level of impact. In contrast to physical transducers like thermistors, photodetectors and movement sensors, chemical and biochemical sensors rely on an intimate contact with the sample, (*e.g.* blood, sweat or tears in the case of WBSN or water or soil in the case of environmental based sensors). These classical chemical sensors and biosensors follow a generic measurement scheme, in which a pre-functionalized surface presents receptor sites that selectively bind a species of interest in a sample.

Since the early breakthroughs in the 1960's and 1970's, which led to the development of a plethora of electrochemical and optochemical diagnostic devices, the vision of reliable and affordable sensors, capable of functioning autonomously over extensive time periods (years) to provide access to continuous streams of real-time data, remains unrealized. This is despite significant investment in research and the many thousands of papers published in the literature. For example, it is over 40 years since the concept of an artificial pancreas was proposed, by combining the glucose electrode with an insulin pump [126]. Even now, there is no chemical sensor/biosensor which can function reliably inside the body for longer than a few days. The root problem remains the impact of biofouling and other processes that rapidly change the response characteristics of the sensor, leading to drift and sensitivity loss. Accordingly, in the past decade scientists have begun to target more accessible media via less invasive means. This is in alignment with the exponential growth of the wearables market, which increasingly seeks to expand the current physical parameters to bring reliable chemical sensing to the wrists of over 3 billion wearers, by 2025 [127].

Likewise, a number of low-cost devices to access molecular information via the analysis of sweat, saliva, interstitial, and ocular fluid have been proposed. At their core, these bodily fluids contain relatively high concentrations of electrolytes, such as sodium, potassium and ammonium salts, in addition to biologically relevant small molecules, such as glucose, lactate, and pyruvate. While the relative concentrations of these compounds in alternative bodily fluids deviate from those found in blood, they offer an accessible means to a wide range of clinically relevant data, which can be collated and analysed to offer wearer-specific models. Several groups have made significant progress towards the realisation of practical platforms for the quantification of electrolytes in sweat in recent years. Integrating ion-selective electrodes with a wearable system capable of harvesting and transporting sweat, watch-type devices have shown impressive ability to harvest sweat and track specific electrolytes in real time [128], as illustrated in Figure 7. Similarly, a wearable electrochemical sensor array has been developed by Javey *et al*. [129]. The resulting fully integrated system, capable of real-time detection of sodium, potassium, glucose, and lactate, is worn as a band on the forehead or arm, transmitting the data to a remote base station.

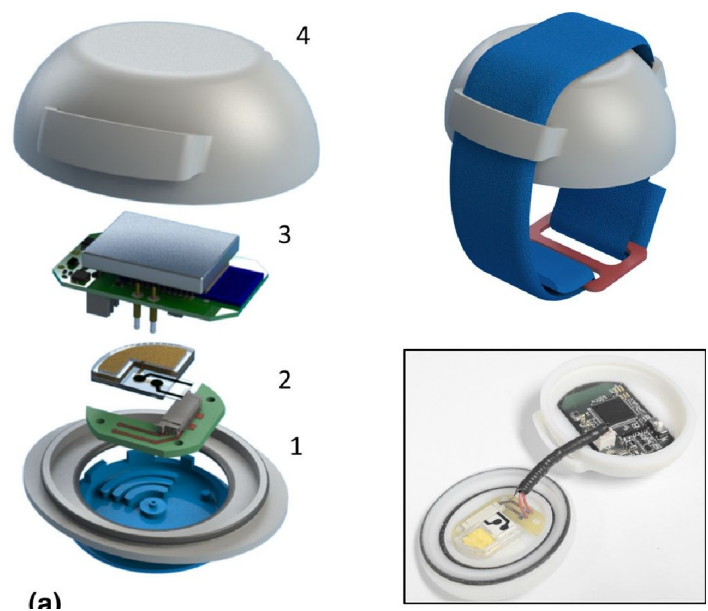

*Figure 7 – SwEatch: watch-sensing platform for sodium analysis in sweat. 1: sweat harvesting device in 3d-printed platform base, 2: fluidic sensing chip, 3: electronic data logger and battery, and 4: 3d-printed upper casing. Reproduced from [128] with permission from the authors.*

Contact lenses provide another means to access a wide range of molecular analytes in a non-invasive manner through the information-rich ocular fluid. Pioneered by Badugu *et al.* [130] nearly 20 years ago, a smart contact lens can monitor ocular glucose through fluorescence changes. The initial design has been further developed to encompass ions such as calcium, sodium, magnesium, and potassium [131]. While the capability of such a device is self-evident, such a restrictive sensing mode may ultimately hamper its application. It took several years for significant inroads into flexible electronics and wireless power transfer to enable a marriage of electrochemical sensing with a conformable contact lens. Demonstrated by Park *et al.* [132] for real-time quantification of glucose in ocular fluid, this platform indicates the potential of combining a reliable, accurate chemical sensing method with integrated power and electronics in a non-invasive approach to access important clinical data.

Although considerably more invasive than sweat or ocular fluid sensing, accurate determination of biomarkers in interstitial fluids has a proven track record. Indeed, the first FDA approved means for non-invasive glucose monitoring, namely the Glucowatch [133], relied on interstitial fluid sampling extracted using reverse iontophoresis with subsequent electrochemical detection. This pioneering development from 2001 offered multiple measurements per hour and provided its wearer with an easy-to-use watch-like interface. Although ultimately hampered by skin-irritation and calibration issues, it nonetheless signified a milestone in minimally invasive glucose measurements. Continuing with this approach, the FDA approved (in 2017) Abbott Freestyle Libre – see Figure 8, enables wirelessly monitoring of blood sugar via analysis of the interstitial fluid. On application, the device punctures the skin and places a 0.5 cm fibre wick through the outer skin barrier, so that interstitial fluid can be sampled and monitored for glucose in real time, for up to two weeks, at which point it is replaced. Data is accessed using a wireless mobile phone-like base station.

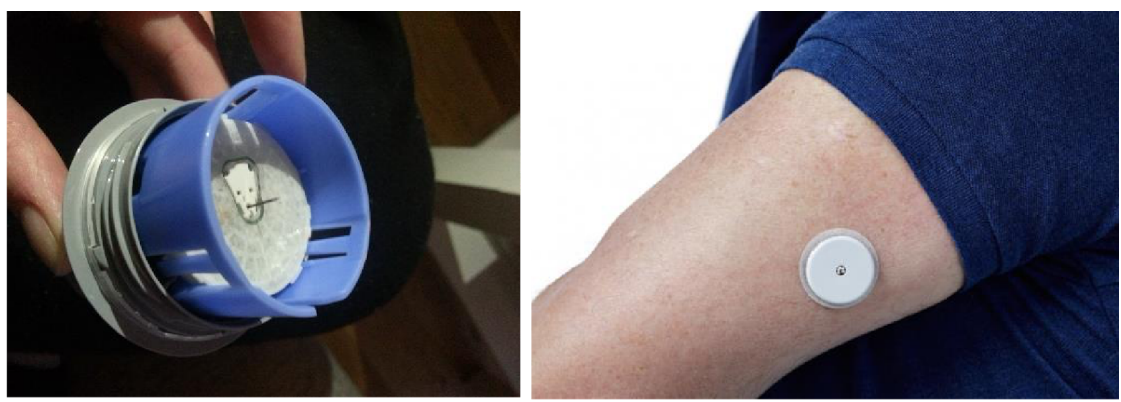

*Figure 8 –The Abbott Libre applicator (left) showing the sampling fibre and (right) the glucose sensing patch on the upper arm after attachment.*

In addition to delivering acceptable analytical performance in the relevant sample media, there are a number of challenges for on-body sensors related to size, rigidity, power, communication, data acquisition, processing, and security [134], which must be overcome before they can realise their full potential and play a pivotal role in applications of IoT in healthcare services.

A similar scenario is faced in the environmental arena, in that (bio)chemical sensing is inherently more expensive and complex than monitoring physical parameters such as temperature, light, depth, or movement. This is strikingly illustrated in the Argo Project, which currently has ca. 3,000-4,000 sensorised

'floats' distributed globally in the oceans, all of which track location, depth, temperature, and salinity. These were originally devised to monitor several core parameters (temperature, pressure, and salinity) and share the data from this global sensor network via satellite communications links, to provide an accurate in-situ picture of the ocean status in real or near real time. Temperature and pressure data are accurate to ±0.002°C, and uncorrected salinity is accurate to ±0.1 psu (can be improved by relatively complex and time-consuming post-acquisition processing). Interestingly, an increasing number of the floats now include 'Biogeochemical' sensors (308, ca. 10%, April 2018) for nitrate (121), chlorophyll (186), oxygen (302) and pH (97), see http://www.argo.ucsd.edu and the maps in Figure 9. Of these, the nitrate measurement is by direct UV absorbance, and chlorophyll by absorbance/fluorescence at spectral regions characteristic of algal chlorophyll; i.e. these are optical measurements rather than conventional chemical sensor measurements. Moreover, it is likely that most, if not all of the pH measurements, are performed using optically responsive dyes rather than the well-known glass electrode. This strikingly demonstrates that chemical sensors are avoided when longer term, reliable and accurate measurements are required from remote locations and hostile environments. It is also striking that more complex chemical and biological measurements (i.e.; that require analysers incorporating reagents, microfluidics etc.) are not included in the Argo project. These autonomous analyser platforms for tracking key parameters like nutrients, concentration of dissolved oxygen (COD), pH, heavy metals, and organics cost typically €15K or more per unit to buy, not including service and consumables charges. For example, the Seabird Electronics dissolved oxygen sensors used in the Argo project cost $60K each [135], and Microlab autonomous environmental phosphate analysers cost ca. €20K per unit; *i.e.* they are far too expensive to use as basic building blocks of larger-scale deployments for IoT applications.

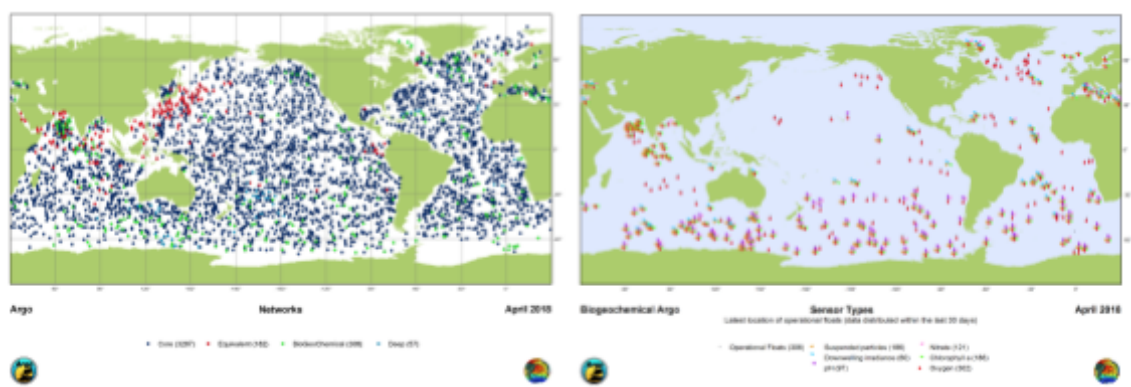

*Figure 9: Distribution of sensorised 'floats' monitoring the status of a variety of ocean parameters that encompass core parameters (temperature, velocity/pressure, salinity) and 'Biogeochemical' sensing (e.g. oxygen, nitrate, chlorophyll, pH).* Reproduced with permission from [136].

Progress towards realizing disruptive improvements is encouraged by competitions organised by environmental agencies such as the *Alliance for Coastal Technologies* (http://www.act-us.info/nutrients-challenge/index.php) who launched in 2015 the 'Global Nutrient Sensor Challenge'. The purpose was to stimulate innovation in the sector, as participants had to deliver nutrient analysers capable of 3-month independent in-situ operation at a maximum unit cost of $5,000. The ACT estimated that the global market for these devices is currently ca. 30,000 units in the USA, and ca. 100,000 units globally (*i.e.* ca. $500 million per year). This is set to expand further as new applications related to nutrient recovery grow in importance (*e.g.* from biodigester units and wastewater treatment plants), driven by the need to meet regulatory targets and business opportunities linked to the rapidly increasing cost of nutrients.

### 4.3 Prospects for Scalable Applications of Chemical Sensors and Biosensors

From the discussion in the previous Section, it is clear that there are substantial markets in personal health and environmental monitoring, and other sectors, for reliable chemical sensors and biosensors that are fit for purpose, with an affordable use model. And while progress has been painfully slow over the past 30-40 years, since the excitement of early promising breakthroughs [137], it is now apparent the beginnings of larger scale use and a tentative move from single-use or centralised facilities, towards real-time continuous measurements at point-of-need. As this trend develops, the range and volume of data collected will rise exponentially, and new types of services will emerge, most likely borrowing ideas and models from existing applications, such as the myriad of products for personal exercise tracking, merged with new tools designed to deal with the more complex behaviour of molecular sensors.

In the health sector, this will support an expansion in the rollout of remote services due to the increasing availability of wearable/implantable diagnostic and autonomous drug delivery platforms that operate in a closed loop control mode with real-time tracking of key biomarkers [138]. Of course, these services will be linked into an overarching personalised health informatics framework that enables healthcare professionals to monitor individual status remotely, and trigger escalations in response if thresholds are breached or a future issue is predicted from data trends.

Likewise, in the environmental sector, unit costs for autonomous chemical analysers for water monitoring remain stubbornly high, constituting a significant barrier to scale up, particularly when coupled with high cost of ownership due to frequent service intervals. This is a frustrating situation, as the tremendous benefits of long term autonomous sensing of key status indicators in the health and the water sectors are clear. For example, low cost, reliable water quality analysers would revolutionise the way drinking, waste and natural waters are monitored. Combining in-situ, real-time water sensing with satellite/flyover remote sensing represents an immensely powerful development, due to the tremendous enhancement of the integrated information content when the global scale of satellite sensing is coupled with the detailed molecular information generated by in-situ deployed sensors and analysers [139]. The scale of data generation from satellite remote sensing is already staggering, already reaching 21.1 PBs (petabytes) by 2015, and continuing to grow exponentially [140]. As Kathryn Sullivan, NOAA[1] Administrator and Under Secretary of Commerce for Oceans and Atmosphere, and former NASA astronaut commented recently; "*NOAA observations alone provide some 20 terabytes every day—twice the data of the Library of Congress' entire print collection*". She also commented on the importance of developing new tools and collaborations to realise the value of the encoded information; "*Just 20 years ago, we were piecing together data points by hand. Five years ago, 90 percent of today's data was yet to be generated. Now we're innovating in the cloud, experiencing Earth with a wider lens and in fresh new ways. NOAA (National Oceanic and Atmospheric Administration) is partnering with Amazon, Microsoft, IBM, Google, and the Open Commons Consortium to tap that potential.*" The global coverage acquired with an increasingly fine spatial resolution, multiple spectral band sensors, and increasing numbers of satellites, is driving this tremendous increase in the scale of data production. However, the benefits are already being realised, for example, through studies that couple highly localised data with large-scale coverage. For example, ecological models have been tested and used to visualise stratification dynamics in 2,368 lakes [141], and daily temperature profiles for almost 11,000 lakes [142]. In the latter case, the models also have been used to predict future water temperatures. In the near future, we are certain to see larger scale in-situ sensing networks, as new technologies emerge that combine lower unit cost with much longer service intervals. New data analysis and visualisation tools will be required to enable these hugely complementary information sources to be combined to maximum effect.

Similarly, it would be a wonderful development if real-time tracking of disease markers could be coupled with smart drug delivery platforms. However, significant fundamental barriers still exist and must be surmounted if this revolution in sensing is to be realised. The most formidable is how to maintain and validate system performance during extended use (typically a minimum of 3 months for water monitoring, and at least several years, preferably 10 years or more for implantables) [143]. In-situ performance validation involves calibration, requiring fluidics and storage of standards/reagents. And all components and solutions must function reliably for the service interval. Progress in water monitoring systems is easier as the interval is shorted and systems are more accessible, with a larger footprint. For implantables, however, the challenges are daunting, and in recent years, researchers have therefore focused on non-invasive/minimally invasive on-body use models as described above [129].

---

[1] https://learn.arcgis.com/en/arcgis-imagery-book/chapter8/# 'Learn-More'
Available from the Environmental Systems Research Institute (ESRI)

If devices capable of meeting the challenge of long-term reliable chemical/biological sensing could be realised, it will represent a keystone breakthrough upon which multiple applications with revolutionary impact on society could be built. It will require informatics systems and tools to process, filter, recognise patterns and events, and communicate with related data from personal, group and, ultimately, to societal levels; from a single location, to global scale. Perhaps when this happens, we will witness at last the emergence of true internet-scale sensing and control via chemical sensors and biosensors, effectively creating a continuum between the molecular and digital worlds [144].

## 5. Current Limitations and Future Outlook

Throughout this review, we adopted an optimistic tone regarding the recent advances and prospects for the use of big data and ML in chemistry and related applications. We believe such optimism is justified by the proliferation of projects – in academia and industry – dedicated to developing artificial intelligence (AI) applications in all possible fields, including chemistry. We take the view that, with massive investment and with so much at stake in the economy of corporations and countries, this ongoing AI-based revolution is unlikely to stop, with many positive prospects for the near future focused on chemistry. Before discussing them, let us concentrate on the limitations and challenges to be faced by chemists and their collaborators in the short and long-term futures within this revolution.

**The state of the art**

In describing the potential and effective use of big data and ML in Section 3, we did not conduct a critical analysis of the examples from the literature. Apart from a few exceptions, we restricted ourselves to highlighting the potential usefulness of these methodologies for several areas of chemistry, with emphasis on materials discovery. Indeed, we considered that identifying limitations (or even deficiencies), in the approaches adopted in specific cases, has limited value in view of the underlying conceptual difficulties to be faced in paving the way for more expressive advances in chemistry supported by big data and ML.

In particular, in the Introduction to this paper, when considering the two types of goals in applying ML, we emphasized that one of them is much harder to reach. Indeed, according to Wallach [4], the two categories of goals are related with "prediction" and "explanation". In the prediction category, observed data are employed to reason about unseen data or missing information. As for the explanation category, the aim is to find plausible explanations for observed data, addressing "why" and "how" questions. As Wallach puts it, *"models for prediction are often intended to replace human interpretation or reasoning, whereas models for explanation are intended to inform or guide human reasoning"*. Regardless of all the recent results attained with ML, as we may recapitulate from Section 3, the examples found in chemistry are not necessarily useful to guide the human reasoning. Despite the quality and usefulness of the results obtained in various studies conducted for "prediction" purposes, the application of ML to generalize and bring entirely new knowledge to chemistry is still to be realized, as confirmed from a quick inspection of the illustrative examples in Section 3.

Furthermore, even in the state of the art of AI and ML, there are no clear hints as to whether and how uses will be possible within the "explanation" category, not only for chemistry but for other areas as well. One may speculate that the answer may result from the convergence of the two big movements mentioned in the Introduction – big data and natural language processing – but the specifics of the solutions are far from established. It is likely, for instance, that a theoretical framework is required to guide experimental design for data collection and to provide conceptual knowledge and effective predictive models [103]. On the other hand, current limitations mean that the full potential of ML in chemistry has yet to be realized.

**Pitfalls**

Critical issues also stem from the combination of ML with big data. That is to say, to which extent having a lot of data to learn from is an actual benefit? Indeed, the overall principle common to big data and ML is that of abandoning physical-chemical simulations that might be infeasible due to the demands of a molecule-to-molecule relationship computation. The workaround is to use formulations and/or high-level material properties to feed ML techniques and teach them to identify patterns; in summary, a trained ML algorithm will work by mapping known physical properties into unknown physical properties. From this perspective, ML may be taken as a curve fitting technique capable of exploring search spaces way bigger than those that could be handled by non-computational approaches. This course of action has actual potential in predicting useful chemical compounds, but most ML techniques, notably DL methods, are not interpretable with respect to their mechanism of action. This fact raises two concerns. First, a chemist learns little or nothing on what caused a given set of outputs, which does not contribute to advancing the field as a principled science. Second, unlike computer vision or speech recognition problems, in which the outputs are directly verifiable, the outputs of ML in chemistry may be biased by a limited training set, or even totally distorted by an ill-defined model. Still, the chemist might take them for granted or, otherwise, take a long time before actually identifying the flaws. Again, such problems are not exclusive to the interplay between chemistry and ML; in fact, they raise growing concerns and motivate ongoing investigations on how to enhance the interpretability of the representations created by these algorithms [145].

Other limitations stem from practical issues in connection with big data. Indeed, such methodologies are not applicable to a variety of problems in chemistry, where the need for collecting and curating high-quality data, and the computational infrastructure required, may pose major challenges. In fact, critical problems have been mentioned in the discussion of sensors and biosensors in Section 4. In order to work efficiently, most learning algorithms require a reasonably large set of known examples. While the cost of collecting, e.g., thousands of images and/and millions of comments in a social network is nearly negligible, experiments in chemistry might demand reagents, enzymes, compounds, combined with significant protocolled labour, and time to observe and annotate the results. As a possible consequence, computational techniques may lack the minimum amount of data necessary to produce trustful models. Rather than saving time and resources, the reverse may happen, leading to increased costs due to erroneous procedures unwittingly applied.

**Challenges and prospects**

In an additional challenge, many societal issues must be addressed to ensure the proper use of big data, including important ethical and privacy preservation questions. Take, for instance, the case of computer-assisted clinical diagnosis, which relies heavily on data from sensors and biosensors. As pointed out by Rodrigues-Jr et al. [40], an important obstacle to the thorough integration of databases is not related to the lack of technology, but, rather, to commitments from individuals and institutions to work together in establishing acceptable procedures for data curation, privacy preservation and to avoid abusive or inadequate usage of medical data. The same holds in chemistry.

In contrast to our approach regarding the uses of big data and ML in chemistry, in reviewing the importance of sensors and biosensors to developing IoT and similar applications, we provided a critical analysis of the advances and limitations in the field. The prospects for the future depend on behavioral issues as well as scientific challenges. We hope that the acquired popularity of big data and ML may raise the awareness of researchers and developers on the importance of how they treat, preserve and analyze their data. We advocate that ML and other computational tools should already be in routine use not only by those working on sensors and biosensors but also in topics not traditionally considered as sensing. The latter include various types of spectroscopy and imaging – which generate massive amounts of data. We believe there is a gap between the wide range of techniques available for data management and analysis and their actual use in the daily practice of many research and development facilities. This may be due to a combination of factors, such as a slow pace of dissemination of novel techniques, lack of a theoretical framework to guide the choice of techniques, and limited availability of accessible and usable implementations. Regardless of the reasons, increased awareness to this issue and more informed use of existing methodologies is an important step to reinforce progress.

As for the prospects for big data and ML for chemistry in the next few years, it is self-evident that considerable advances can be attained by extending the many examples seen in Section 3. One may, for instance, envisage searching for drugs and drug targets by harnessing the whole body of medical literature, as a complement to applying those almost entirely-chemistry oriented approaches. An example of such a system is already being tested with a version of the Watson supercomputer [146], which deals with text in scientific papers and patents, besides considering pharmacological, chemical and genomics data. The rationale behind the Watson approach is to establish connections in millions of text pages and in a huge database of molecular properties. Cases of enhanced prediction with clinical chemistry data were illustrated by Richardson et al. [7], who employ ML methods with large datasets for diseases such as hepatitis B and C. Nonetheless, the major goal of reaching truly intelligent systems to solve problems in chemistry beyond those of classification types will require the convergence of big data and ML within schemes that allow one to interpret and explain the results. This is a major challenge not only for chemistry, but for any area of science and technology.

**Acknowledgments:**

This work received support from Brazilian agencies FAPESP (2013/14262-7, 2018/17620-5), CNPq and CAPES. L.F and D.D acknowledge support from Science Foundation Ireland from the Insight Centre for Data Analytics, grant SFI/12/RC/2289. The authors are also grateful to Dr. Don Pierson, from the Evolutionary Biology Centre, Uppsala, Sweden, for helpful discussions on the scale of satellite remote sensed data.

# References

(1) Chen, C. L. P.; Zhang, C.-Y. Data-Intensive Applications, Challenges, Techniques and Technologies: A Survey on Big Data. *Inf. Sci. (Ny)*. **2014**, *275*, 314–347.

(2) Lusher, S. J.; McGuire, R.; van Schaik, R. C.; Nicholson, C. D.; de Vlieg, J. Data-Driven Medicinal Chemistry in the Era of Big dData. *Drug Discov. Today* **2014**, *19* (7), 859–868.

(3) dos Santos, L. B.; Júnior, E. A. C.; Jr., O. N. O.; Amancio, D. R.; Mansur, L. L.; Aluísio, S. M. Enriching Complex Networks with Word Embeddings for Detecting Mild Cognitive Impairment from Speech Transcripts. In *Proceedings of the 55th Annual Meeting of the Association for Computational Linguistics, ACL*; 2017; Vol. 1, pp 1284–1296.

(4) Wallach, H. Computational Social Science ≠ Computer Science + Social Data. *Commun. ACM* **2018**, *61* (3), 42–44.

(5) Akimushkin, C.; Amancio, D. R.; Oliveira, O. N. On the Role of Words in the Network Structure of Texts: Application to Authorship Attribution. *Phys. A Stat. Mech. its Appl.* **2018**, *495*, 49–58.

(6) Alpaydin, E. *Introduction to Machine Learning*, 2nd ed.; The MIT Press, 2010.

(7) Richardson, A.; Signor, B. M.; Lidbury, B. A.; Badrick, T. Clinical Chemistry in Higher Dimensions: Machine-Learning and Enhanced Prediction from Routine Clinical Chemistry Data. *Clin. Biochem.* **2016**, *49* (16–17), 1213–1220.

(8) Gantz, J.; Reinsel, D. Extracting Value from Chaos. *IDC IView* **2011**, No. 1142, 1–12. Replace by: https://www.forbes.com/sites/bernardmarr/2018/05/21/how-much-data-do-we-create-every-day-the-mind-blowing-stats-everyone-should-read/#1f7f393960ba

(9) Alvarez-Moreno, M.; de Graaf, C.; Lopez, N.; Maseras, F.; Poblet, J. M.; Bo, C. Managing the Computational Chemistry Big Data Problem: The IoChem-BD Platform. *J. Chem. Inf. Model.* **2015**, *55* (1), 95–103.

(10) Xie, Y.-S.; Kumar, D.; Bodduri, V. D. V.; Tarani, P. S.; Zhao, B.-X.; Miao, J.-Y.; Jang, K.; Shin, D.-S. Microwave-Assisted Parallel Synthesis of Benzofuran-2-Carboxamide Derivatives Bearing Anti-Inflammatory, Analgesic and Antipyretic Agents. *Tetrahedron Lett.* **2014**, *55* (17), 2796–2800.

(11) Gao, H.; Korn, J. M.; Ferretti, S.; Monahan, J. E.; Wang, Y.; Singh, M.; Zhang, C.; Schnell, C.; Yang, G.; Zhang, Y.; et al. High-Throughput Screening Using Patient-Derived Tumor Xenografts to Predict Clinical Trial Drug Response. *Nat. Med.* **2015**, *21* (11), 1318–1325.

(12) Gilson, M. K.; Liu, T.; Baitaluk, M.; Nicola, G.; Hwang, L.; Chong, J. BindingDB in 2015: A Public Database for Medicinal Chemistry, Computational Chemistry and Systems Pharmacology. *Nucleic Acids Res.* **2016**, *44* (D1), D1045--53.

(13) Schneider, N.; Lowe, D. M.; Sayle, R. A.; Tarselli, M. A.; Landrum, G. A. Big Data from Pharmaceutical Patents: A Computational Analysis of Medicinal Chemists' Bread and Butter. *J. Med. Chem.* **2016**, *59* (9), 4385–4402.

(14) Tetko, I. V; Engkvist, O.; Koch, U.; Reymond, J.-L.; Chen, H. BIGCHEM: Challenges and

Opportunities for Big Data Analysis in Chemistry. *Mol. Inform.* **2016**, *35* (11–12), 615–621.

(15) Kelleher, J. D.; Namee, B. Mac; D'Arcy, A. *Fundamentals of Machine Learning for Predictive Data Analytics: Algorithms, Worked Examples, and Case Studies*; The MIT Press, 2015.

(16) Verma, J.; Khedkar, V. M.; Coutinho, E. C. 3D-QSAR in Drug Design--a Review. *Curr. Top. Med. Chem.* **2010**, *10* (1), 95–115.

(17) LeCun, Y.; Bengio, Y.; Hinton, G. Deep Learning. *Nature* **2015**, *521* (7553), 436–444.

(18) Krizhevsky, A.; Sutskever, I.; Hinton, G. E. ImageNet Classification with Deep Convolutional Neural Networks. In *Proceedings of the 25th International Conference on Neural Information Processing Systems*; NIPS'12; Curran Associates Inc.: USA, 2012; Vol. 1, pp 1097–1105.

(19) Lecun, Y.; Bottou, L.; Bengio, Y.; Haffner, P. Gradient-Based Learning Applied to Document Recognition. *Proc. IEEE* **1998**, *86* (11), 2278–2324.

(20) Bahrampour, S.; Ramakrishnan, N.; Schott, L.; Shah, M. Comparative Study of Deep Learning Software Frameworks. **2016**.

(21) Goh, G. B.; Hodas, N. O.; Vishnu, A. Deep Learning for Computational Chemistry. *J. Comput. Chem.* **2017**, *38* (16), 1291–1307.

(22) Hartnett, M.; Diamond, D.; Barker, P. G. Neural Network Based Recognition of Flow Injection Patterns. *Analyst* **1993**, *118* (4), 347–354.

(23) Lindsay, R. K.; Buchanan, B. G.; Feigenbaum, E. A.; Lederberg, J. DENDRAL: A Case Study of the First Expert System for Scientific Hypothesis Formation. *Artif. Intell.* **1993**, *61* (2), 209–261.

(24) Szymkuc, S.; Gajewska, E. P.; Klucznik, T.; Molga, K.; Dittwald, P.; Startek, M.; Bajczyk, M.; Grzybowski, B. A. Computer-Assisted Synthetic Planning: The End of the Beginning. *Angew. Chem. Int. Ed. Engl.* **2016**, *55* (20), 5904–5937.

(25) Le, T. C.; Winkler, D. A. Discovery and Optimization of Materials Using Evolutionary Approaches. *Chem. Rev.* **2016**, *116* (10), 6107–6132.

(26) Katritzky, A. R.; Lobanov, V. S.; Karelson, M. QSPR: The Correlation and Quantitative Prediction of Chemical and Physical Properties from Structure. *Chem. Soc. Rev.* **1995**, *24* (4), 279–287.

(27) Le, T.; Epa, V. C.; Burden, F. R.; Winkler, D. A. Quantitative Structure-Property Relationship Modeling of Diverse Materials Properties. *Chem. Rev.* **2012**, *112* (5), 2889–2919.

(28) Kalinin, S. V; Sumpter, B. G.; Archibald, R. K. Big-Deep-Smart Data in Imaging for Guiding Materials Design. *Nat. Mater.* **2015**, *14* (10), 973–980.

(29) Ward, L.; Wolverton, C. Atomistic Calculations and Materials Informatics: A Review. *Curr. Opin. Solid State Mater. Sci.* **2017**, *21* (3), 167–176.

(30) Breneman, C. M.; Brinson, L. C.; Schadler, L. S.; Natarajan, B.; Krein, M.; Wu, K.; Morkowchuk, L.; Li, Y.; Deng, H.; Xu, H. Stalking the Materials Genome: A Data-Driven Approach to the Virtual Design of Nanostructured Polymers. *Adv. Funct. Mater.* **2013**, *23* (46), 5746–5752.

(31) Jain, A.; Ong, S. P.; Hautier, G.; Chen, W.; Richards, W. D.; Dacek, S.; Cholia, S.; Gunter, D.; Skinner, D.; Ceder, G.; et al. Commentary: The Materials Project: A Materials Genome Approach to Accelerating Materials Innovation. *APL Mater.* **2013**, *1* (1), 11002.

(32) Subcommittee on the Materials Genome Initiative. The First Five Years of the Materials Genome Initiative: Accomplishments and Technical Highlights. National Science and Technology Council - Committee on Technology. Executive Office of the President of the United States. mgi.gov/sites/default/files/documents/mgi-accomplishments-at-5-years-august-2016.pdf (Accessed on 04/22/2018) 2016, pp 1–9.

(33) Nakata, M.; Shimazaki, T. PubChemQC Project: A Large-Scale First-Principles Electronic Structure Database for Data-Driven Chemistry. *J. Chem. Inf. Model.* **2017**, *57* (6), 1300–1308.

(34) Tibbetts, K. W. M.; Li, R.; Pelczer, I.; Rabitz, H. Discovering Predictive Rules of Chemistry from Property Landscapes. *Chem. Phys. Lett.* **2013**, *572*, 1–12.

(35) Wolf, D.; Buyevskaya, O. V; Baerns, M. An Evolutionary Approach in the Combinatorial Selection and Optimization of Catalytic Materials. *Appl. Catal. A Gen.* **2000**, *200* (1), 63–77.

(36) Bulut, M.; Gevers, L. E. M.; Paul, J. S.; Vankelecom, I. F. J.; Jacobs, P. A. Directed Development of High-Performance Membranes via High-Throughput and Combinatorial Strategies. *J. Comb. Chem.* **2006**, *8* (2), 168–173.

(37) Corey, E. J.; Wipke, W. T. Computer-Assisted Design of Complex Organic Syntheses. *Science* **1969**, *166* (3902), 178–192.

(38) Coley, C. W.; Barzilay, R.; Jaakkola, T. S.; Green, W. H.; Jensen, K. F. Prediction of Organic Reaction Outcomes Using Machine Learning. *ACS Cent. Sci.* **2017**, *3* (5), 434–443.

(39) Liao, S.-H. Expert System Methodologies and Applications—a Decade Review from 1995 to 2004. *Expert Syst. Appl.* **2005**, *28* (1), 93–103.

(40) Rodrigues-Jr, J. F.; Paulovich, F. V; de Oliveira, M. C.; de Oliveira, O. N. J. On the Convergence of Nanotechnology and Big Data Analysis for Computer-Aided Diagnosis. *Nanomedicine (Lond).* **2016**, *11* (8), 959–982.

(41) Segler, M. H. S.; Waller, M. P. Modelling Chemical Reasoning to Predict and Invent Reactions. *Chem. – A Eur. J.* **2017**, *23* (25), 6118–6128.

(42) Schwaller, P.; Gaudin, T.; Lanyi, D.; Bekas, C.; Laino, T. "Found in Translation": Predicting Outcomes of Complex Organic Chemistry Reactions Using Neural Sequence-to-Sequence Models. **2017**.

(43) Cadeddu, A.; Wylie, E. K.; Jurczak, J.; Wampler-Doty, M.; Grzybowski, B. A. Organic Chemistry as a Language and the Implications of Chemical Linguistics for Structural and Retrosynthetic Analyses. *Angew. Chem. Int. Ed. Engl.* **2014**, *53* (31), 8108–8112.

(44) Lowe, D. M. Extraction of Chemical Structures and Reactions from the Literature, PhD Thesis at University of Cambridge, 2012.

(45) Weininger, D. SMILES, a Chemical Language and Information System. 1. Introduction to

Methodology and Encoding Rules. *J. Chem. Inf. Comput. Sci.* **1988**, *28* (1), 31–36.

(46) Jin, W.; Coley, C. W.; Barzilay, R.; Jaakkola, T. S. Predicting Organic Reaction Outcomes with Weisfeiler-Lehman Network. In *Advances in Neural Information Processing Systems (NIPS), Long Beach, CA, USA*; 2017; pp 2604–2613.

(47) Gómez-Bombarelli, R.; Wei, J. N.; Duvenaud, D.; Hernández-Lobato, J. M.; Sánchez-Lengeling, B.; Sheberla, D.; Aguilera-Iparraguirre, J.; Hirzel, T. D.; Adams, R. P.; Aspuru-Guzik, A. Automatic Chemical Design Using a Data-Driven Continuous Representation of Molecules. *ACS Cent. Sci.* **2018**, *4* (2), 268–276.

(48) Segler, M. H. S.; Waller, M. P. Neural-Symbolic Machine Learning for Retrosynthesis and Reaction Prediction. *Chemistry* **2017**, *23* (25), 5966–5971.

(49) Ramakrishnan, R.; Dral, P. O.; Rupp, M.; von Lilienfeld, O. A. Big Data Meets Quantum Chemistry Approximations: The Delta-Machine Learning Approach. *J. Chem. Theory Comput.* **2015**, *11* (5), 2087–2096.

(50) Dral, P. O.; von Lilienfeld, O. A.; Thiel, W. Machine Learning of Parameters for Accurate Semiempirical Quantum Chemical Calculations. *J. Chem. Theory Comput.* **2015**, *11* (5), 2120–2125.

(51) Lopez-Bezanilla, A.; von Lilienfeld, O. A. Modeling Electronic Quantum Transport with Machine Learning. *Phys. Rev. B* **2014**, *89* (23), 235411.

(52) Kolb, B.; Lentz, L. C.; Kolpak, A. M. Discovering Charge Density Functionals and Structure-Property Relationships with PROPhet: A General Framework for Coupling Machine Learning and First-Principles Methods. *Sci. Rep.* **2017**, *7* (1192), 1–9.

(53) Finkelmann, A. R.; Goller, A. H.; Schneider, G. Site of Metabolism Prediction Based on Ab Initio Derived Atom Representations. *ChemMedChem* **2017**, *12* (8), 606–612.

(54) Pereira, F.; Xiao, K.; Latino, D. A. R. S.; Wu, C.; Zhang, Q.; Aires-de-Sousa, J. Machine Learning Methods to Predict Density Functional Theory B3LYP Energies of HOMO and LUMO Orbitals. *J. Chem. Inf. Model.* **2017**, *57* (1), 11–21.

(55) Deringer, V. L.; Csányi, G.; Proserpio, D. M. Extracting Crystal Chemistry from Amorphous Carbon Structures. *ChemPhysChem* **2017**, *18* (8), 873–877.

(56) Ramakrishnan, R.; Hartmann, M.; Tapavicza, E.; von Lilienfeld, O. A. Electronic Spectra from TDDFT and Machine Learning in Chemical Space. *J. Chem. Phys.* **2015**, *143* (8), 84111.

(57) Dral, P. O.; Owens, A.; Yurchenko, S. N.; Thiel, W. Structure-Based Sampling and Self-Correcting Machine Learning for Accurate Calculations of Potential Energy Surfaces and Vibrational Levels. *J. Chem. Phys.* **2017**, *146* (24), 244108.

(58) Janet, J. P.; Kulik, H. J. Predicting Electronic Structure Properties of Transition Metal Complexes with Neural Networks. *Chem. Sci.* **2017**, *8* (7), 5137–5152.

(59) Schymanski, E. L.; Ruttkies, C.; Krauss, M.; Brouard, C.; Kind, T.; Duhrkop, K.; Allen, F.; Vaniya, A.; Verdegem, D.; Bocker, S.; et al. Critical Assessment of Small Molecule Identification 2016: Automated Methods. *J. Cheminform.* **2017**, *9* (1), 22.

(60) Rupp, M.; Tkatchenko, A.; Muller, K.-R.; von Lilienfeld, O. A. Fast and Accurate Modeling of Molecular Atomization Energies with Machine Learning. *Phys. Rev. Lett.* **2012**, *108* (5), 58301.

(61) Montavon, G.; Rupp, M.; Gobre, V.; Vazquez-Mayagoitia, A.; Hansen, K.; Tkatchenko, A.; Müller, K.-R.; Anatole von Lilienfeld, O. Machine Learning of Molecular Electronic Properties in Chemical Compound Space. *New J. Phys.* **2013**, *15* (9), 95003.

(62) Pyzer‐Knapp, E. O.; Li, K.; Aspuru‐Guzik, A. Learning from the Harvard Clean Energy Project: The Use of Neural Networks to Accelerate Materials Discovery. *Adv. Funct. Mater.* **2015**, *25* (41), 6495–6502.

(63) Mohimani, H.; Gurevich, A.; Mikheenko, A.; Garg, N.; Nothias, L.-F.; Ninomiya, A.; Takada, K.; Dorrestein, P. C.; Pevzner, P. A. Dereplication of Peptidic Natural Products through Database Search of Mass Spectra. *Nat. Chem. Biol.* **2017**, *13* (1), 30–37.

(64) Pires, D. E. V; Blundell, T. L.; Ascher, D. B. PkCSM: Predicting Small-Molecule Pharmacokinetic and Toxicity Properties Using Graph-Based Signatures. *J. Med. Chem.* **2015**, *58* (9), 4066–4072.

(65) Cheng, F.; Li, W.; Zhou, Y.; Shen, J.; Wu, Z.; Liu, G.; Lee, P. W.; Tang, Y. AdmetSAR: A Comprehensive Source and Free Tool for Assessment of Chemical ADMET Properties. *J. Chem. Inf. Model.* **2012**, *52* (11), 3099–3105.

(66) Cheng, F.; Ikenaga, Y.; Zhou, Y.; Yu, Y.; Li, W.; Shen, J.; Du, Z.; Chen, L.; Xu, C.; Liu, G.; et al. In Silico Assessment of Chemical Biodegradability. *J. Chem. Inf. Model.* **2012**, *52* (3), 655–669.

(67) Cheng, F.; Shen, J.; Yu, Y.; Li, W.; Liu, G.; Lee, P. W.; Tang, Y. In Silico Prediction of Tetrahymena Pyriformis Toxicity for Diverse Industrial Chemicals with Substructure Pattern Recognition and Machine Learning Methods. *Chemosphere* **2011**, *82* (11), 1636–1643.

(68) Cheng, F.; Yu, Y.; Shen, J.; Yang, L.; Li, W.; Liu, G.; Lee, P. W.; Tang, Y. Classification of Cytochrome P450 Inhibitors and Noninhibitors Using Combined Classifiers. *J. Chem. Inf. Model.* **2011**, *51* (5), 996–1011.

(69) Cheng, F.; Yu, Y.; Zhou, Y.; Shen, Z.; Xiao, W.; Liu, G.; Li, W.; Lee, P. W.; Tang, Y. Insights into Molecular Basis of Cytochrome P450 Inhibitory Promiscuity of Compounds. *J. Chem. Inf. Model.* **2011**, *51* (10), 2482–2495.

(70) Shen, J.; Cheng, F.; Xu, Y.; Li, W.; Tang, Y. Estimation of ADME Properties with Substructure Pattern Recognition. *J. Chem. Inf. Model.* **2010**, *50* (6), 1034–1041.

(71) Broccatelli, F.; Carosati, E.; Neri, A.; Frosini, M.; Goracci, L.; Oprea, T. I.; Cruciani, G. A Novel Approach for Predicting P-Glycoprotein (ABCB1) Inhibition Using Molecular Interaction Fields. *J. Med. Chem.* **2011**, *54* (6), 1740–1751.

(72) Karelson, M.; Lobanov, V. S.; Katritzky, A. R. Quantum-Chemical Descriptors in QSAR/QSPR Studies. *Chem. Rev.* **1996**, *96* (3), 1027–1044.

(73) Gramatica, P. Principles of QSAR Models Validation: Internal and External. *QSAR Comb. Sci.* **2007**, *26* (5), 694–701.

(74) Tropsha, A. Best Practices for QSAR Model Development, Validation, and Exploitation. *Mol. Inform.*

**2010**, *29* (6–7), 476–488.

(75) Kaggle Team. Deep Learning How I Did It: Merck 1st Place Interview. http://blog.kaggle.com/2012/11/01/deep-learning-how-i-did-it-merck-1st-place-interview/ (Accessed on 04/22/2018) 2012, p 1.

(76) Mauri, A.; Consonni, V.; Pavan, M.; Todeschini, R. Dragon Software: An Easy Approach to Molecular Descriptor Calculations. *Match Commun. Math. Comput. Chem.* **2006**, *56* (2), 237–248.

(77) Mayr, A.; Klambauer, G.; Unterthiner, T.; Hochreiter, S. DeepTox: Toxicity Prediction Using Deep Learning. *Front. Environ. Sci.* **2016**, *3*, 80.

(78) Wallach, I.; Dzamba, M.; Heifets, A. AtomNet: A Deep Convolutional Neural Network for Bioactivity Prediction in Structure-Based Drug Discovery. *CoRR* **2015**, *1510.02855*.

(79) Xu, Y.; Dai, Z.; Chen, F.; Gao, S.; Pei, J.; Lai, L. Deep Learning for Drug-Induced Liver Injury. *J. Chem. Inf. Model.* **2015**, *55* (10), 2085–2093.

(80) Ekins, S.; Spektor, A. C.; Clark, A. M.; Dole, K.; Bunin, B. A. Collaborative Drug Discovery for More Medicines for Tuberculosis (MM4TB). *Drug Discov. Today* **2017**, *22* (3), 555–565.

(81) Schneider, G.; Schneider, P. Macromolecular Target Prediction by Self-Organizing Feature Maps. *Expert Opin. Drug Discov.* **2017**, *12* (3), 271–277.

(82) Müller, A. T.; Kaymaz, A. C.; Gabernet, G.; Posselt, G.; Wessler, S.; Hiss, J. A.; Schneider, G. Sparse Neural Network Models of Antimicrobial Peptide‐Activity Relationships. *Mol. Inform.* **2016**, *35* (11–12), 606–614.

(83) Porto, W. F.; Pires, A. S.; Franco, O. L. Antimicrobial Activity Predictors Benchmarking Analysis Using Shuffled and Designed Synthetic Peptides. *J. Theor. Biol.* **2017**, *426*, 96–103.

(84) Gawehn, E.; Hiss, J. A.; Schneider, G. Deep Learning in Drug Discovery. *Mol. Inform.* **2016**, *35* (1), 3–14.

(85) Mistry, P.; Neagu, D.; Trundle, P. R.; Vessey, J. D. Using Random Forest and Decision Tree Models for a New Vehicle Prediction Approach in Computational Toxicology. *Soft Comput.* **2016**, *20* (8), 2967–2979.

(86) Ekins, S. The Next Era: Deep Learning in Pharmaceutical Research. *Pharm. Res.* **2016**, *33* (11), 2594–2603.

(87) Korotcov, A.; Tkachenko, V.; Russo, D. P.; Ekins, S. Comparison of Deep Learning with Multiple Machine Learning Methods and Metrics Using Diverse Drug Discovery Data Sets. *Mol. Pharm.* **2017**, *14* (12), 4462–4475.

(88) Marks, D.; Hopf, T.; Sander, C. Protein Structure Prediction from Sequence Variation. *Nat. Biotechnol.* **2012**, *30*, 1072–1080.

(89) Di Lena, P.; Nagata, K.; Baldi, P. Deep Architectures for Protein Contact Map Prediction. *Bioinformatics* **2012**, *28* (19), 2449–2457.

(90) Fox, N. K.; Brenner, S. E.; Chandonia, J.-M. SCOPe: Structural Classification of Proteins--Extended,

Integrating SCOP and ASTRAL Data and Classification of New Structures. *Nucleic Acids Res.* **2014**, *42* (Database issue), D304--9.

(91) Wang, F.; Liu, S.; Shu, L.; Tao, X.-M. Low-Dimensional Carbon Based Sensors and Sensing Network for Wearable Health and Environmental Monitoring. *Carbon N. Y.* **2017**, *121*, 353–367.

(92) Oliveira Jr, O. N.; Iost, R. M.; Siqueira, J. R. J.; Crespilho, F. N.; Caseli, L. Nanomaterials for Diagnosis: Challenges and Applications in Smart Devices Based on Molecular Recognition. *ACS Appl. Mater. Interfaces* **2014**, *6* (17), 14745–14766.

(93) Uçar, A.; Özalp, R. Efficient Android Electronic Nose Design for Recognition and Perception of Fruit Odors Using Kernel Extreme Learning Machines. *Chemom. Intell. Lab. Syst.* **2017**, *166*, 69–80.

(94) Zhang, L.; Liu, Y.; Deng, P. Odor Recognition in Multiple E-Nose Systems With Cross-Domain Discriminative Subspace Learning. *IEEE Trans. Instrum. Meas.* **2017**, *66* (7), 1679–1692.

(95) Tomazzoli, M. M.; Pai Neto, R. D.; Moresco, R.; Westphal, L.; Zeggio, A. R. S.; Specht, L.; Costa, C.; Rocha, M.; Maraschin, M. Discrimination of Brazilian Propolis According to the Seasoning Using Chemometrics and Machine Learning Based on UV-Vis Scanning Data. *J. Integr. Bioinform.* **2015**, *12* (4), 279.

(96) Wetterich, C. B.; Felipe de Oliveira Neves, R.; Belasque, J.; Ehsani, R.; Marcassa, L. G. Detection of Huanglongbing in Florida Using Fluorescence Imaging Spectroscopy and Machine-Learning Methods. *Appl. Opt.* **2017**, *56* (1), 15–23.

(97) Mutlu, A. Y.; Kilic, V.; Ozdemir, G. K.; Bayram, A.; Horzum, N.; Solmaz, M. E. Smartphone-Based Colorimetric Detection via Machine Learning. *Analyst* **2017**, *142* (13), 2434–2441.

(98) Paulovich, F. V; Moraes, M. L.; Maki, R. M.; Ferreira, M.; Oliveira Jr., O. N.; de Oliveira, M. C. F. Information Visualization Techniques for Sensing and Biosensing. *Analyst* **2011**, *136* (7), 1344–1350.

(99) Oliveira, O. N.; Pavinatto, F. J.; Constantino, C. J. L.; Paulovich, F. V; de Oliveira, M. C. F. Information Visualization to Enhance Sensitivity and Selectivity in Biosensing. *Biointerphases* **2012**, *7* (1–4), 1–15.

(100) Inselberg, A. The Plane with Parallel Coordinates. *Vis. Comput.* **1985**, *1* (2), 69–91.

(101) Thapa, A.; Soares, A. C.; Soares, J. C.; Awan, I. T.; Volpati, D.; Melendez, M. E.; Fregnani, J. H. T. G.; Carvalho, A. L.; Oliveira, O. N. J. Carbon Nanotube Matrix for Highly Sensitive Biosensors To Detect Pancreatic Cancer Biomarker CA19-9. *ACS Appl. Mater. Interfaces* **2017**, *9* (31), 25878–25886.

(102) Moraes, M. L.; Petri, L.; Oliveira, V.; Olivati, C. A.; de Oliveira, M. C. F.; Paulovich, F. V; Oliveira, O. N.; Ferreira, M. Detection of Glucose and Triglycerides Using Information Visualization Methods to Process Impedance Spectroscopy Data. *Sensors Actuators B Chem.* **2012**, *166–167*, 231–238.

(103) Aileni, R. M. Healthcare Predictive Model Based on Big Data Fusion from Biomedical Sensors. In *ELearning Vision 2020!*; 2016; Vol. 1, pp 328–333.

(104) McCarthy, J. F.; Marx, K. A.; Hoffman, P. E.; Gee, A. G.; O'Neil, P.; Ujwal, M. L.; Hotchkiss, J. Applications of Machine Learning and High-Dimensional Visualization in Cancer Detection,

Diagnosis, and Management. *Ann. N. Y. Acad. Sci.* **2004**, *1020*, 239–262.

(105) Legin, A.; Rudnitskaya, A.; Lvova, L.; Vlasov, Y.; Natale, C. Di; D'Amico, A. Evaluation of Italian Wine by the Electronic Tongue: Recognition, Quantitative Analysis and Correlation with Human Sensory Perception. *Anal. Chim. Acta* **2003**, *484* (1), 33–44.

(106) Daikuzono, C. M.; Shimizu, F. M.; Manzoli, A.; Riul, A.; Piazzetta, M. H. O.; Gobbi, A. L.; Correa, D. S.; Paulovich, F. V; Oliveira, O. N. Information Visualization and Feature Selection Methods Applied to Detect Gliadin in Gluten-Containing Foodstuff with a Microfluidic Electronic Tongue. *ACS Appl. Mater. Interfaces* **2017**, *9* (23), 19646–19652.

(107) Ferreira, E. J.; Pereira, R. C. T.; Delbem, A. C. B.; Oliveira, O. N.; Mattoso, L. H. C. Random Subspace Method for Analysing Coffee with Electronic Tongue. *Electron. Lett.* **2007**, *43* (21), 1138–1139.

(108) Byrne, R.; Diamond, D. Chemo/Bio-Sensor Networks. *Nat. Mater.* **2006**, *5*, 421.

(109) IOT Tech Center, TNLIST, Tsinghua. GreenOrbs. http://www.greenorbs.org/ (Accessed on 05/04/2018).

(110) Beckwith, R.; Teibel, D.; Bowen, P. Report from the Field: Results from an Agricultural Wireless Sensor Network. In *29th Annual IEEE International Conference on Local Computer Networks*; 2004; pp 471–478.

(111) Burrell, J.; Brooke, T.; Beckwith, R. Vineyard Computing: Sensor Networks in Agricultural Production. *IEEE Pervasive Comput.* **2004**, *3* (1), 38–45.

(112) Morais, R.; Fernandes, M. A.; Matos, S. G.; Serôdio, C.; Ferreira, P. J. S. G.; Reis, M. J. C. S. A ZigBee Multi-Powered Wireless Acquisition Device for Remote Sensing Applications in Precision Viticulture. *Comput. Electron. Agric.* **2008**, *62* (2), 94–106.

(113) MicroStrain, Inc. Shelburne Vineyard Relies on Wireless Sensors and the Cloud to Monitor Its Vines. http://www.microstrain.com/support/news/shelburne-vineyard-relies-wireless-sensors-and-cloud-monitor-its-vines (Accessed on 05/04/2018) 2012.

(114) Werner-Allen, G.; Lorincz, K.; Ruiz, M.; Marcillo, O.; Johnson, J.; Lees, J.; Welsh, M. Deploying a Wireless Sensor Network on an Active Volcano. *IEEE Internet Comput.* **2006**, *10* (2), 18–25.

(115) Park, D.-H.; Park, J.-W. Wireless Sensor Network-Based Greenhouse Environment Monitoring and Automatic Control System for Dew Condensation Prevention. *Sensors (Basel).* **2011**, *11* (4), 3640–3651.

(116) Mekki, M.; Abdallah, O.; Amin, M. B. M.; Eltayeb, M.; Abdalfatah, T.; Babiker, A. Greenhouse Monitoring and Control System Based on Wireless Sensor Network. In *International Conference on Computing, Control, Networking, Electronics and Embedded Systems Engineering (ICCNEEE)*; 2015; pp 384–387.

(117) Cardell-Oliver, R.; Kranz, M.; Smettem, K.; Mayer, K. A Reactive Soil Moisture Sensor Network: Design and Field Evaluation. *Int. J. Distrib. Sens. Networks* **2005**, *1* (2), 149–162.

(118) Diamond, D.; Lau, K. T.; Brady, S.; Cleary, J. Integration of Analytical Measurements and Wireless

Communications-Current Issues and Future Strategies. *Talanta* **2008**, *75* (3), 606–612.

(119) Larios, D. F.; Barbancho, J.; Sevillano, J. L.; Rodriguez, G.; Molina, F. J.; Gasull, V. G.; Mora-Merchan, J. M.; Leon, C. Five Years of Designing Wireless Sensor Networks in the Donana Biological Reserve (Spain): An Applications Approach. *Sensors (Basel).* **2013**, *13* (9), 12044–12069.

(120) Martonosi, M. Embedded Systems in the Wild: ZebraNet Software, Hardware, and Deployment Experiences. *SIGPLAN Not.* **2006**, *41* (7), 1.

(121) Xu, G.; Shen, W.; Wang, X. Applications of Wireless Sensor Networks in Marine Environment Monitoring: A Survey. *Sensors (Basel).* **2014**, *14* (9), 16932–16954.

(122) Johnson, K. S.; Needoba, J. A.; Riser, S. C.; Showers, W. J. Chemical Sensor Networks for the Aquatic Environment. *Chem. Rev.* **2007**, *107* (2), 623–640.

(123) Aqeel-Ur-Rehman; Abbasi, A. Z.; Islam, N.; Shaikh, Z. A. A Review of Wireless Sensors and Networks' Applications in Agriculture. *Comput. Stand. Interfaces* **2014**, *36* (2), 263–270.

(124) Hollander, R. Two-Thirds of the World's Population Are Now Connected by Mobile Devices. Business Insider UK. Sep. 2017. http://uk.businessinsider.com/world-population-mobile-devices-2017-9 (Accessed on 05/04/2018).

(125) Jakicic, J. M.; Davis, K. K.; Rogers, R. J.; King, W. C.; Marcus, M. D.; Helsel, D.; Rickman, A. D.; Wahed, A. S.; Belle, S. H. Effect of Wearable Technology Combined With a Lifestyle Intervention on Long-Term Weight Loss: The IDEA Randomized Clinical Trial. *JAMA* **2016**, *316* (11), 1161–1171.

(126) Albisser, A. M.; Leibel, B. S.; Ewart, T. G.; Davidovac, Z.; Botz, C. K.; Zingg, W.; Schipper, H.; Gander, R. Clinical Control of Diabetes by the Artificial Pancreas. *Diabetes* **1974**, *23* (5), 397–404.

(127) Hayward, J.; Pugh, D.; Chansin, G. Wearable Sensors 2018-2028: Technologies, Markets & Players. IDTechEx. http://www.idtechex.com/research/reports/wearable-sensors-2018-2028-technologies-markets-and-players-000555.asp (Accessed on 04/06/2018) 2017, pp 1–292.

(128) Tom, G.; Conor, O.; Margaret, M.; Giusy, M.; Stephen, B.; G., W. G.; Florin, S.; Niamh, O.; Paddy, W.; Dermot, D. "SWEATCH": A Wearable Platform for Harvesting and Analysing Sweat Sodium Content. *Electroanalysis* **2016**, *28* (6), 1283–1289.

(129) Gao, W.; Emaminejad, S.; Nyein, H. Y. Y.; Challa, S.; Chen, K.; Peck, A.; Fahad, H. M.; Ota, H.; Shiraki, H.; Kiriya, D.; et al. Fully Integrated Wearable Sensor Arrays for Multiplexed in Situ Perspiration Analysis. *Nature* **2016**, *529*, 509–514.

(130) Badugu, R.; Lakowicz, J. R.; Geddes, C. D. Noninvasive Continuous Monitoring of Physiological Glucose Using a Monosaccharide-Sensing Contact Lens. *Anal. Chem.* **2004**, *76* (3), 610–618.

(131) Badugu, R.; Jeng, B. H.; Reece, E. A.; Lakowicz, J. R. Contact Lens to Measure Individual Ion Concentrations in Tears and Applications to Dry Eye Disease. *Anal. Biochem.* **2018**, *542*, 84–94.

(132) Park, J.; Kim, J.; Kim, S.-Y.; Cheong, W. H.; Jang, J.; Park, Y.-G.; Na, K.; Kim, Y.-T.; Heo, J. H.; Lee, C. Y.; et al. Soft, Smart Contact Lenses with Integrations of Wireless Circuits, Glucose Sensors, and Displays. *Sci. Adv.* **2018**, *4* (1).

(133) Tierney, M. J.; Tamada, J. A.; Potts, R. O.; Jovanovic, L.; Garg, S. Clinical Evaluation of the GlucoWatch Biographer: A Continual, Non-Invasive Glucose Monitor for Patients with Diabetes. *Biosens. Bioelectron.* **2001**, *16* (9–12), 621–629.

(134) Bandodkar, A. J.; Jeerapan, I.; Wang, J. Wearable Chemical Sensors: Present Challenges and Future Prospects. *ACS Sensors* **2016**, *1* (5), 464–482.

(135) Argo. Part of the Integrated Global Observation Strategy. http://www.argo.ucsd.edu/ (Accessed on 06/04/2018).

(136) Argo. Argo Float Data and Metadata from Global Data Assembly Centre (Argo GDAC). http://dx.doi.org/10.17882/42182 (Accessed on 06/04/2018) 2000.

(137) H Garrett DeYoung. Biosensors - The Mating of Biology and Electronics. *High Technol.* **1983**, No. 11, 41–49.

(138) Patra, D.; Sengupta, S.; Duan, W.; Zhang, H.; Pavlick, R.; Sen, A. Intelligent, Self-Powered, Drug Delivery Systems. *Nanoscale* **2013**, *5* (4), 1273–1283.

(139) McCaul, M.; Barland, J.; Cleary, J.; Cahalane, C.; McCarthy, T.; Diamond, D. Combining Remote Temperature Sensing with In-Situ Sensing to Track Marine/Freshwater Mixing Dynamics. *Sensors* **2016**, *16* (9), 1402–1418.

(140) Fan, J.; Yan, J.; Ma, Y.; Wang, L. Big Data Integration in Remote Sensing across a Distributed Metadata-Based Spatial Infrastructure. *Remote Sens.* **2017**, *10* (2), 7.

(141) Read, J. S.; Winslow, L. A.; Hansen, G. J. A.; Van Den Hoek, J.; Hanson, P. C.; Bruce, L. C.; Markfort, C. D. Simulating 2368 Temperate Lakes Reveals Weak Coherence in Stratification Phenology. *Ecol. Modell.* **2014**, *291*, 142–150.

(142) Winslow, L. A.; Hansen, G. J. A.; Read, J. S.; Notaro, M. Large-Scale Modeled Contemporary and Future Water Temperature Estimates for 10774 Midwestern U.S. Lakes. *Sci. Data* **2017**, *4*, 170053.

(143) Coleman, S.; Florea, L.; Diamond, D. Chemical Sensing with Autonomous Devices in Remote Locations - Why Is It so Difficult and How Do We Deliver Revolutionary Improvements in Performance. *Irish Chem. News* **2016**, No. 1, February, 13–23.

(144) Diamond, D. Internet-Scale Sensing. *Anal. Chem.* **2004**, *76* (15), 278A–286A.

(145) Yosinski, J.; Clune, J.; Nguyen, A. M.; Fuchs, T. J.; Lipson, H. Understanding Neural Networks Through Deep Visualization. *CoRR* **2015**, *abs/1506.0*.

(146) Chen, Y.; Argentinis, J. D. E.; Weber, G. IBM Watson: How Cognitive Computing Can Be Applied to Big Data Challenges in Life Sciences Research. *Clin. Ther.* **2016**, *38* (4), 688–701.

(147) Ramakrishnan, Raghunathan, Dral, Pavlo O, Rupp,Matthias, and Von Lilienfeld, O Anatole.Quan-tum chemistry structures and properties of 134 kilomolecules. Scientific data, 1, 2014.

(148) Curtarolo, S. et al. AFLOWLIB.ORG: A distributed materials properties repository from

high-throughput ab initio calculations. Comp. Mater. Sci. **2012,** 58, 227–235.

(149) Smith, J. S.; Isayev, O.; Roitberg, A. E. Data Descriptor: ANI-1, A Data Set of 20 Million Calculated off-Equilibrium Conformations for Organic Molecules. Sci, **2017**.

(150) Ian Goodfellow, Yoshua Bengio, Aaron Courville. Deep Learning (Adaptive Computation and Machine Learning series). The MIT Press, **2016**.

(151) Whitley D, Sutton AM. Genetic algorithms a survey of models and methods. In: Rozenberg G, Bäck T, Kok JN (eds) Handbook of natural computing. Springer, Berlin, **2012,** 637–671.

(152) Wojciech Paszkowicz. Genetic Algorithms, a Nature-Inspired Tool: Survey of Applications in Materials Science and Related Fields, Materials and Manufacturing Processes, **2009,** 24:2**,** 174-197.